\documentclass[final,5p,times,twocolumn]{elsarticle}%
\usepackage{graphicx}
\usepackage{epstopdf}
\usepackage{bm}
\usepackage{amssymb}
\usepackage{epsfig}
\usepackage{subfigure}
\usepackage{amsmath}
\usepackage{amsfonts}
\usepackage{color}
\usepackage{multirow}
\usepackage[noend]{algpseudocode}
\usepackage{natbib}
\setcitestyle{authoryear,round}
\usepackage[colorlinks, 
linkcolor = blue, 
anchorcolor = blue, citecolor = blue]{hyperref}
\usepackage{algorithmicx,algorithm}%
\providecommand{\U}[1]{\protect\rule{.1in}{.1in}}
\journal{Automatica}
\newtheorem{definition}{\rm\textbf{Definition}}
\newtheorem{theorem}{\rm\textbf{Theorem}}

\newtheorem{assump}{\rm\textbf{Assumption}}
\newtheorem{remark}{\rm\textbf{Remark}}
\newtheorem{problem}{\textbf{Problem}}

\begin{document}
\begin{frontmatter}
\title{Bridging the Gap between Optimal Trajectory Planning and Safety-Critical Control with Applications to Autonomous Vehicles\tnoteref{thanks} %
\tnotetext[thanks]{This work was supported in part by NSF under grants ECCS-1931600, DMS-1664644, CNS-1645681, IIS-1723995, and CPS-1446151, by ARPA-E's NEXTCAR program under grant DE-AR0000796, by AFOSR under grant FA9550-19-1-0158, and by the MathWorks.}}
\author[BU]{Wei Xiao} \ead{xiaowei@bu.edu*}    
\author[BU]{Christos G. Cassandras} \ead{cgc@bu.edu}               
\author[BU]{Calin A. Belta} \ead{cbelta@bu.edu}
\address[BU]{Division of Systems Engineering, Boston University, Brookline, MA, USA}
\begin{abstract}
We address the problem of optimizing the performance of a dynamic system while satisfying hard safety constraints at all times.
Implementing an optimal control solution is limited by the computational cost required to derive it in real time,
especially when constraints become active, as well as the need to rely on simple linear dynamics,
simple objective functions, and ignoring noise.
The recently proposed Control Barrier Function (CBF) method may be used for safety-critical control at the expense of
sub-optimal performance.
In this paper, we develop a real-time control framework that combines optimal trajectories generated through optimal
control with the computationally efficient CBF method providing safety guarantees. We use Hamiltonian analysis to
obtain a tractable optimal solution for a linear or linearized system, then employ High Order CBFs (HOCBFs) and
Control Lyapunov Functions (CLFs) to account for constraints with arbitrary relative degrees and to track the
optimal state, respectively. We further show how to deal with noise in arbitrary relative degree systems.
The proposed framework is then applied to the optimal traffic merging problem for Connected and Automated
Vehicles (CAVs) where the objective is to jointly minimize the travel time and energy consumption of each CAV
subject to speed, acceleration, and speed-dependent safety constraints. In addition, when considering more complex
objective functions, nonlinear dynamics and passenger comfort requirements for which
analytical optimal control solutions are unavailable, we adapt the HOCBF method to such problems.
Simulation examples are included to compare the performance of the proposed framework to optimal solutions
(when available) and to a baseline provided by human-driven vehicles with results showing significant improvements
in all metrics.
\end{abstract}
\end{frontmatter}

\section{INTRODUCTION}

Optimizing a cost function associated with the operation of a dynamical system
while also satisfying hard safety constraints at all times is a fundamental
and challenging problem. The challenge is even greater when stabilizing some
system state variables to desired values is an additional requirement. With
the growing role of autonomy, the importance of these problems has also grown
and one now frequently encounters them in the operation of autonomous vehicles
in robotics and traffic networks. These applications provide the main
motivation for the control framework presented in this paper.

Optimal control problems with safety-critical constraints can be solved
through standard methods \citet{Bryson1969}, \citet{Ansari2016}, with
applications found in robotics and autonomous vehicles in traffic networks
\citet{Chitour2012}, \citet{Mita2001}, \citet{Malikopoulos2018}, \citet{Wei2019CDC}.
However, analytical solutions are only possible for simple system dynamics and
constraints. Moreover, the computational complexity for deriving such
solutions significantly increases as one or more constraints become active and
it grows as a power function of the number of constraints. This fact limits
the use of optimal control methods for autonomous systems when solutions need
to be derived and executed on line. Additional factors which further limit the
real-time use of these methods include the presence of noise in the dynamics,
model inaccuracies, environmental perturbations, and communication delays in
the information exchange among system components. Thus, there is a gap between
optimal control solutions (which represent a lower bound for the optimal
achievable cost) and the execution of controllers aiming to achieve such
solutions under realistic operational conditions.

In order to bridge this gap and obtain real-time controls for safety-critical
problems, Model Predictive Control (MPC) \citet{Garcia1989}, \citet{Mayne2014}, \citet{Bemporad2002} has been widely used to approximate optimal control
solutions. Whether linear or nonlinear MPC methods are used, a
time-discretized predictive model is needed and a receding horizon control
problem is formulated and solved at all discretized receding time instants
taking into account all safety constraints involved. Nonetheless, the
computational cost significantly increases with the model nonlinearity and the
time horizon over which a problem is solved.

An alternative approach which has the potential to avoid the drawbacks above
is based on the use of Control Barrier Function (CBFs). Barrier functions are
Lyapunov-like functions \citet{Wieland2007} whose use can be traced back to
optimization theory \citet{Boyd2004}. More recently, they have been employed in
verification and control, e.g., to prove set invariance \citet{Aubin2009}, \citet{Prajna2007}, \citet{Wisniewski2013}, and for multi-objective control
\citet{Panagou2013}. CBFs are extensions of barrier functions for control
systems \citet{Ames2017} and have been recently generalized to consider
arbitrary relative degree constraints in \citet{Nguyen2016}, \citet{Xiao2019}. It
has also been shown that CBFs can be combined with Control Lyapunov Functions
(CLFs) \citet{Sontag1983}, \citet{Freeman1996}, \citet{Aaron2012} to form
constrained quadratic programs (QPs) \citet{Galloway2013} for nonlinear control
systems that are affine in controls. The main advantages of CBF-based control
compared to MPC lie in the fact that $(i)$ feasible state sets under CBF-based
control possess a forward invariance guarantee property, $(ii)$ The QPs
involved at every time step can be solved in real time, as long as each QP is
feasible, and $(iii)$ the method is easier to adapt when handling nonlinear
systems with complex constraints.

{The contribution of this paper is to synthesize controllers that
\emph{combine} the optimal control and the CBF methods aiming for both
optimality and guaranteed safety in real-time control. The key idea is to
first generate trajectories by solving a tractable optimal control problem and
then seek to track these trajectories using a controller which simultaneously
ensures that all state and control constraints are satisfied at all times.
This is accomplished in two steps. The first step is to solve a constrained
optimal control problem. Given a set of initial conditions, it is usually
possible to derive simple conditions under which it can be shown that no
constraint becomes active. In this case, executing the unconstrained optimal
control solution becomes a relatively simple tracking problem. Otherwise, we
can still often derive an optimal control solution consisting of both
unconstrained and constrained arcs. However, such derivations may not always
be feasible in real time. Either way, using the best possible analytical
solution within reasonable real-time computational constraints (possibly just
the unconstrained solution), this step leads to a reference control
$u_{ref}(t)$, $t\in\lbrack0,T]$. The second step is then to use High Order
CBFs (HOCBFs) \citet{Xiao2019} to account for constraints with arbitrary
relative degrees, and define a sequence of QPs whose goal is to
\emph{optimally track }$u_{ref}(t)$ at each discrete time step over $[0,T]$.
In this step, we can allow noise in the system dynamics and include
nonlinearities which were ignored in the original optimal control solution.
The resulting controller is termed \emph{Optimal control with Control Barrier
Functions (OCBF)}. We will show that using an OCBF controller we can achieve
near-optimal performance relative to the one under optimal control while
guaranteeing constraint satisfaction under more general dynamics and the
presence of disturbances that the original optimal control solution cannot
capture.}

The OCBF idea was used in our recent work \citet{Wei2019itsc} to address the
merging problem for Connected Automated Vehicles (CAVs) in traffic networks.
This is one of the most challenging problems within a transportation system in
terms of safety, congestion, and energy consumption, in addition to being a
source of stress for many drivers \citet{Schrank2015}, \citet{Tideman2007}, \citet{Waard2009}. More broadly, advances in transportation system
technologies and the emergence of CAVs have the potential to drastically
improve a transportation network's performance by better assisting drivers in
making decisions, ultimately reducing energy consumption, air pollution,
congestion and accidents. Early efforts exploiting the benefit of CAVs were
proposed in \citet{Levine1966}, \citet{Varaiya1993}. In terms of optimal
trajectory planning, a number of centralized and decentralized merging control
mechanisms have been proposed \citet{Milanes2012}, \citet{Tideman2007}, \citet{Raravi2007}, \citet{Scarinci2014}. In the case of
decentralized control, all computation is performed on board each vehicle and
shared only with a small number of other vehicles which are affected by it.
The objectives specified for optimal control problems may target the
minimization of acceleration as in \citet{Torres2015} or the maximization of
passenger comfort (measured as the acceleration derivative or jerk) as in
\citet{Ntousakis2016}, \citet{Rathgeber2015}. MPC techniques are employed as an
alternative, primarily to account for additional constraints and to compensate
for disturbances by re-evaluating optimal actions \citet{Cao2015}, \citet{Mukai2017}, \citet{Ntousakis2016}. As an alternative to MPC, CBF methods
were used in \citet{Wei2019} where a decentralized optimal control problem with
explicit analytical solutions for each CAV was derived.

In this paper, we generalize the OCBF controller introduced in
\citet{Wei2019itsc} that only works for relative degree one constraints to
allow constraints with relative degree greater than one and also allow for
noise in the system dynamics. We consider optimal control problems with
constraints of arbitrary relative degrees which are handled by using HOCBFs.
We will show that by using HOCBFs we can incorporate complex objective
functions, nonlinear dynamics, and comfort requirements which otherwise
prohibit even unconstrained optimal control solutions from being derived. This
also allows us to study the trade-off between travel time, energy consumption,
and comfort. Extensive simulations have been conducted to demonstrate the
effectiveness of the proposed framework for the traffic merging problem
relative to other approaches.

The paper is structured as follows. In Section \ref{sec:pre}, we provide
definitions and results on the HOCBF method. We formulate a general
constrained optimal control problem and develop its OCBF solution in Sections
\ref{sec:prob_general} and \ref{sec:solution}, respectively. {As an
application of the OCBF framework, in Section \ref{sec:prob} we present the
traffic merging process model and formulate the optimal merging control
problem including all safety, state and control constraints that must be
satisfied at all times.} In Section \ref{sec:ocbf}, the optimal solution for
the merging problem is reviewed for the unconstrained as well as the
constrained cases and the OCBF method is applied to it. We provide simulation
examples and performance comparisons with human-driven vehicles in Section
\ref{sec:simulation} and conclude with Section \ref{sec:conclude}.

\section{PRELIMINARIES}

\label{sec:pre}

Consider an affine control system of the form
\begin{equation}
\dot{\bm{x}}=f(\bm x)+g(\bm x)\bm u \label{eqn:affine}%
\end{equation}
where $\bm x\in X\subset\mathbb{R}^{n}$, $f:\mathbb{R}^{n}\rightarrow
\mathbb{R}^{n}$ and $g:\mathbb{R}^{n}\rightarrow\mathbb{R}^{n\times q}$ are
globally Lipschitz, and $\bm u\in U\subset\mathbb{R}^{q}$ ($U$ denotes the
control constraint set). Solutions $\bm x(t)$ of (\ref{eqn:affine}), starting
at $\bm x(0)$, $t\geq0$, are forward complete. The control constraint set $U$
is defined as (the inequality is interpreted componentwise, $\bm u_{min},\bm
u_{max}\in\mathbb{R}^{q}$):
\begin{equation}
U:=\{\bm u\in\mathbb{R}^{q}:\bm u_{min}\leq\bm u\leq\bm u_{max}\}.
\label{eqn:control}%
\end{equation}

\begin{definition}
\label{def:classk} (\textit{Class $\mathcal{K}$ function} \citet{Khalil2002}) A
continuous function $\alpha:[0,a)\rightarrow[0,\infty), a > 0$ is said to
belong to class $\mathcal{K}$ if it is strictly increasing and $\alpha(0)=0$.
\end{definition}

\begin{definition}
\label{def:forwardinv} A set $C\subset\mathbb{R}^{n}$ is forward invariant for
system (\ref{eqn:affine}) if its solutions starting at any $\bm x(0) \in C$
satisfy $\bm x(t)\in C, \forall t\geq0$.
\end{definition}

\begin{definition}
\label{def:relative} (\textit{Relative degree}) The relative degree of a
(sufficiently many times) differentiable function $b:\mathbb{R}^{n}%
\rightarrow\mathbb{R}$ with respect to system (\ref{eqn:affine}) is the number
of times it is differentiated along the dynamics (\ref{eqn:affine}) until the
control $\bm u$ explicitly shows in the corresponding derivative.
\end{definition}

In this paper, the function $b$ is used to define a constraint $b(\bm x)\geq
0$. Therefore, we will also refer to the relative degree of $b$ as the
relative degree of the constraint. For a constraint $b(\bm x)\geq0$ with
relative degree $m$, $b:\mathbb{R}^{n}\rightarrow\mathbb{R}$, and $\psi
_{0}(\bm x):=b(\bm x)$, we define a sequence of functions $\psi_{i}%
:\mathbb{R}^{n}\rightarrow\mathbb{R},i\in\{1,\dots,m\}$:
\begin{equation}
\begin{aligned} \psi_i(\bm x) := \dot \psi_{i-1}(\bm x) + \alpha_i(\psi_{i-1}(\bm x)),\;\;\;i\in\{1,\dots,m\}, \end{aligned} \label{eqn:functions}%
\end{equation}
where $\alpha_{i}(\cdot),i\in\{1,\dots,m\}$ denotes a $(m-i)^{th}$ order
differentiable class $\mathcal{K}$ function. We further define a sequence of
sets $C_{i},i\in\{1,\dots,m\}$ associated with (\ref{eqn:functions}) in the
form:
\begin{equation}
\begin{aligned} C_i := \{\bm x \in \mathbb{R}^n: \psi_{i-1}(\bm x) \geq 0\}, \;\;\;i\in\{1,\dots,m\}. \end{aligned} \label{eqn:sets}%
\end{equation}

\begin{definition}
\label{def:hocbf} (\textit{High Order Control Barrier Function (HOCBF)}
\citet{Xiao2019}) Let $C_{1},\dots,C_{m}$ be defined by (\ref{eqn:sets}) and
$\psi_{1}(\bm x),\dots,\psi_{m}(\bm x)$ be defined by (\ref{eqn:functions}). A
function $b:\mathbb{R}^{n}\rightarrow\mathbb{R}$ is a high order control
barrier function (HOCBF) of relative degree $m$ for system (\ref{eqn:affine})
if there exist $(m-i)^{th}$ order differentiable class $\mathcal{K}$ functions
$\alpha_{i},i\in\{1,\dots,m-1\}$ and a class $\mathcal{K}$ function
$\alpha_{m}$ such that
\begin{equation}
\begin{aligned} \sup_{\bm u\in U}[L_f^{m}b(\bm x) \!+\! L_gL_f^{m-1}b(\bm x)\bm u \!+\! S(b(\bm x)) \!+\! \alpha_m(\psi_{m-1}(\bm x))] \geq 0 \end{aligned} \label{eqn:constraint}%
\end{equation}
for all $\bm x\in C_{1}\cap,\dots,\cap C_{m}$. In (\ref{eqn:constraint}),
$L_{f}^{m}$ ($L_{g}$) denotes Lie derivatives along $f$ ($g$) $m$ (one) times,
and $S(\cdot)$ denotes the remaining Lie derivatives along $f$ with degree
less than or equal to $m-1$.
\end{definition}

The HOCBF constraints in (\ref{eqn:constraint}) may sometimes conflict with
the control constraints in (\ref{eqn:control}), which can limit the existence
of feasible solutions for the optimal control problem that we will formulate
later. In order to minimize this effect, the penalty method \citet{Xiao2019}
replaces $\alpha_{i}(\psi_{i-1}(\bm x))$ by $p_{i}\cdot\alpha_{i}(\psi
_{i-1}(\bm x)), \forall i\in\{1,\dots,m\}$, where $p_{i}>0$ is a
multiplicative penalty factor which can be tuned appropriately.

\begin{theorem}
\label{thm:hocbf} (\citet{Xiao2019}) Given a HOCBF $b(\bm x)$ from Def.
\ref{def:hocbf} with the associated sets $C_{1},\dots,C_{m}$ defined by
(\ref{eqn:sets}), if $\bm x(0)\in C_{1}\cap,\dots,\cap C_{m}$, then any
Lipschitz continuous controller $\bm u(t)\in U$ that satisfies
(\ref{eqn:constraint}), $\forall t\geq0$ renders $C_{1}\cap,\dots,\cap C_{m}$
forward invariant for system (\ref{eqn:affine}).
\end{theorem}

The HOCBF is a general form of the relative degree one CBF \citet{Ames2017},
\citet{Glotfelter2017} (i.e., setting $m=1$ reduces the HOCBF to the common CBF
form in \citet{Ames2017}, \citet{Glotfelter2017}). The exponential CBF
\citet{Nguyen2016} is a special case of the HOCBF.

\begin{definition}
\label{def:clf} (\textit{Control Lyapunov function (CLF)} \citet{Aaron2012}) A
continuously differentiable function $V: \mathbb{R}^{n}\rightarrow\mathbb{R}$
is a globally and exponentially stabilizing control Lyapunov function (CLF)
for system (\ref{eqn:affine}) if there exist constants $c_{1} >0, c_{2}>0,
c_{3}>0$ such that
\begin{equation}
c_{1}||\bm x||^{2} \leq V(\bm x) \leq c_{2} ||\bm x||^{2}%
\end{equation}
\begin{equation}
\label{eqn:clfp}\inf_{u\in U} \lbrack L_{f}V(\bm x)+L_{g}V(\bm x) \bm u +
c_{3}V(\bm x)\rbrack\leq0.
\end{equation}
for $\forall\bm x\in\mathbb{R}^{n}$.
\end{definition}

\begin{theorem}
\label{thm:clf} (\citet{Aaron2012}) Given an exponentially stabilizing CLF $V$
as in Def. \ref{def:clf}, any Lipschitz continuous controller $\bm u(t)\in U$
that satisfies (\ref{eqn:clfp}), $\forall t\geq0$ exponentially stabilizes
system (\ref{eqn:affine}) to the origin.
\end{theorem}

Note that (\ref{eqn:clfp}) can be relaxed by replacing $0$ by a relaxation
variable $\delta\geq0$ at its right-hand side which can be subsequently
minimized \citet{Aaron2012}.

Many existing works \citet{Ames2017},\citet{Lindemann2019},\citet{Nguyen2016}
combine CBFs for systems with relative degree one with quadratic costs to form
optimization problems. Time is discretized and an optimization problem with
constraints given by the CBFs (inequalities of the form (\ref{eqn:constraint}%
)) is solved at each time step. If convergence to a state is desired, then a
CLF constraint of the form (\ref{eqn:clfp}) is added. Note that these
constraints are linear in control since the state value is fixed at the
beginning of the interval, therefore, each optimization problem is a quadratic
program (QP). The optimal control obtained by solving each QP is applied at
the current time step and held constant for the whole interval. The state is
updated using dynamics (\ref{eqn:affine}), and the procedure is repeated.
Replacing CBFs by HOCBFs allows us to handle constraints with arbitrary
relative degree \citet{Xiao2019}.

\section{PROBLEM FORMULATION AND APPROACH}

\label{sec:prob_general}

$\mathbf{Objective}$: (Cost minimization) Consider an optimal control problem
for system (\ref{eqn:affine}) with the cost defined as:
\begin{equation}
J=\int_{t_{0}}^{t_{f}}\left[  \beta+\mathcal{C}%
(\bm x,\bm u,t)\right]  dt, \label{eqn:gcost}%
\end{equation}
where $t_{0},t_{f}$ denote the initial and final times, respectively, and
$\mathcal{C}:\mathbb{R}^{n}\times\mathbb{R}^{q}\times\lbrack t_{0}%
,t_{f}]\rightarrow\mathbb{R}^{+}$ is a cost function. The parameter $\beta
\geq0$ is used to capture a trade-off between the minimization of the time
interval $(t_{f}-t_{0})$ and the operational cost $\mathcal{C}(\bm x,\bm
u,t)$. The terminal time $t_{f}$ is constrained as follows:

\textbf{Terminal state constraint}: The state of system (\ref{eqn:affine}) is
constrained to reach a point $\bar{\bm X}\in{X}$, i.e.,
\begin{equation}
\bm x(t_{f})=\bar{\bm X}, \label{eqn:target}%
\end{equation}
Note that $t_{f}$ is generally free (unspecified).

\textbf{Constraint 1} (Safety constraints): Let $S_{o}$ denote an index set
for a set of safety constraints. System (\ref{eqn:affine}) should always
satisfy
\begin{equation}
b_{j}(\bm x(t))\geq0,\text{ \ }\forall t\in\lbrack t_{0},t_{f}].
\label{eqn:safetycons}%
\end{equation}
where each $b_{j}:\mathbb{R}^{n}\rightarrow\mathbb{R},j\in S_{o}$ is
continuously differentiable.

\textbf{Constraint 2} (Control constraints): These are provided by the control
constraint set in (\ref{eqn:control}).

\textbf{Constraint 3} (State constraints): System (\ref{eqn:affine}) should
always satisfy the state constraints (componentwise):
\begin{equation}
\bm x_{\min}\leq\bm x(t)\leq\bm x_{\max},\forall t\in\lbrack t_{0},t_{f}]
\label{eqn:state}%
\end{equation}
where $\bm x_{\min}\in\mathbb{R}^{n}$ and $\bm x_{\max}\in\mathbb{R}^{n}$.
Note that we distinguish the state constraints from the safety constraints in
(\ref{eqn:safetycons}) {since the latter are viewed as hard, while
the former usually capture system capability limitations that can be relaxed
to improve the problem feasibility}; for example, in traffic networks vehicles
are constrained by upper and lower speed limits.

\begin{problem}
\label{prob:general} Find a control policy for system (\ref{eqn:affine}) such
that the cost (\ref{eqn:gcost}) is minimized, constraints
(\ref{eqn:safetycons}),(\ref{eqn:state}) and (\ref{eqn:control}) are strictly
satisfied, and deviations $||\bm x(t_{f})-\bar{\bm X}||^{2}$ from the terminal
state constraint (\ref{eqn:target}) are minimized.
\end{problem}

The cost in (\ref{eqn:gcost}) can be properly normalized by defining
$\beta:=\frac{\alpha\sup_{\bm x\in X,\bm u\in U,\tau\in\lbrack t_{0},t_{f}%
]}\mathcal{C}(\bm x,\bm u,\tau)}{(1-\alpha)}$ where $\alpha\in\lbrack0,1)$ and
then multiplying (\ref{eqn:gcost}) by $\frac{\alpha}{\beta}$. Thus, we
construct a convex combination as follows:
\begin{equation}
\setlength{\abovedisplayskip}{2pt}\setlength{\belowdisplayskip}{2pt}\begin{aligned}J= \int_{t_0}^{t_f}\left(\alpha + \frac{(1-\alpha)\mathcal{C}(\bm x, \bm u, t)}{\sup_{\bm x\in X, \bm u\in U, \tau\in [t_0,t_f]}\mathcal{C}(\bm x, \bm u, \tau)}\right)dt \end{aligned}.
\label{eqn:cost}%
\end{equation}
If $\alpha=1$, then we solve (\ref{eqn:gcost}) as a minimum time problem. The
normalized cost (\ref{eqn:cost}) facilitates a trade-off analysis between the
two metrics. However, we will use the simpler cost expression (\ref{eqn:gcost}%
) throughout this paper. Thus, we can take $\beta\geq0$ as a weight factor
that can be adjusted to penalize time relative to the cost $\mathcal{C}%
(\bm x,\bm u,t)$ in (\ref{eqn:gcost}).

\textbf{Approach:} \textit{Step 1}: We use Hamiltonian analysis to obtain an
optimal control $\bm u^{\ast}(t)$ and optimal state $\bm x^{\ast}(t)$,
$t\in\lbrack t_{0},{t}_{f}]$ for the cost (\ref{eqn:gcost}) and system
(\ref{eqn:affine}), under the terminal state constraint (\ref{eqn:target}),
the safety constraints (\ref{eqn:safetycons}), and the control and state
constraints (\ref{eqn:control}), (\ref{eqn:state}). In order to get an
analytical optimal solution, we may linearize or simplify the dynamics
(\ref{eqn:affine}).

\textit{Step 2}: There are usually unmodelled dynamics and measurement noise
in (\ref{eqn:affine}). Thus, we consider a modified version of system
(\ref{eqn:affine}) to denote the real dynamics:
\begin{equation}
\dot{\bm{x}}=f(\bm x)+g(\bm x)\bm u+\bm w, \label{eqn:affine_noi}%
\end{equation}
where $\bm w\in\mathbb{R}^{n}$ denotes a vector of random processes in an
appropriate probability space {intended to capture disturbances for
which a precise model is generally unknown. We consider $\bm x$ as a measured
state which includes the effects of such unmodelled dynamics and measurement
noise and which can be used in what follows.} Allowing for the noisy dynamics
(\ref{eqn:affine_noi}), we set $\bm u_{ref}(t)=\bm u^{\ast}(t)$ (more
generally, $\bm u_{ref}(t)=h(\bm u^{\ast}(t),\bm x^{\ast}(t),\bm
x(t)),h:\mathbb{R}^{q}\times\mathbb{R}^{n}\times\mathbb{R}^{n}\rightarrow
\mathbb{R}^{q}$) and use the HOCBF method to track the optimal control as a
reference, i.e.,
\begin{equation}
\setlength{\abovedisplayskip}{1pt}\setlength{\belowdisplayskip}{1pt}\min
_{\bm u(t)}\int_{t_{0}}^{t_{f}}||\bm u(t)-\bm u_{ref}(t)||^{2}dt \label{eqn:t}%
\end{equation}
subject to $(i)$ the HOCBF constraints (\ref{eqn:constraint}) corresponding to
the safety constraints (\ref{eqn:safetycons}), $(ii)$ the state constraints
(\ref{eqn:state}), and $(iii)$ the control constraints (\ref{eqn:control}). In
order to better track the optimal state $\bm x^{\ast}(t)$ and minimize the
deviation $||\bm x(t_{f})-\bar{\bm X}||^{2}$ from the terminal state
constraint, we define a CLF $V(\bm x-\bm x^{\ast})$. Thus, the cost
(\ref{eqn:t}) is also subject to the corresponding CLF constraint
(\ref{eqn:clfp}). The resulting problem can then be solved by the approach
described at the end of Sec. \ref{sec:pre}.

\section{FROM PLANNING TO EXECUTION}

\label{sec:solution}

In this section, we describe how to solve Problem \ref{prob:general} combining
optimality with safety guarantees.

\subsection{Optimal Trajectory Planning\label{sec:optimalsolution}}

\label{sec:oc} Let us consider a {properly linearized version} of
(\ref{eqn:affine_noi}) without the noise $\bm w$:
\begin{equation}
\dot{\bm{x}}=A\bm x+B\bm u, \label{eqn:linear}%
\end{equation}
where $\bm x=(x_{1},\dots,x_{n}),\bm u=(u_{1},\dots,u_{q}),A\in\mathbb{R}%
^{n\times n},B\in\mathbb{R}^{n\times q}$.

Let $\bm\lambda(t)$ be the costate vector corresponding to the state $\bm x$
in (\ref{eqn:linear}) and $\bm b(\bm
x)$ denote the vector obtained by concatenating all $b_{j}(\bm x),j\in S_{o}$.
The Hamiltonian with the state constraints, control constraints and safety
constraints adjoined (omitting time arguments for simplicity) is
\begin{equation}
\begin{aligned} H(\bm x,\bm\lambda, \bm u) = \mathcal{C}(\bm x, \bm u, t) + \bm\lambda^T(A\bm x + B\bm u) + \bm\mu_a^T(\bm u - \bm u_{\max}) \\+ \bm\mu_b^T(\bm u_{\min} - \bm u) + \bm\mu_c^T(\bm x - \bm x_{\max}) + \bm\mu_d^T(\bm x_{\min} - \bm x) \\-\bm \mu_{e}^T\bm b(\bm x) + \beta \end{aligned} \label{eqn:H1}%
\end{equation}
The components of the Lagrange multiplier vectors $\bm\mu_{a},\bm\mu
_{b},\bm\mu_{c},\bm\mu_{d},\bm\mu_{e}$ are positive when the constraints are
active and become 0 when the constraints are strict.

First, we assume all the constraints (\ref{eqn:control}),
(\ref{eqn:safetycons}), (\ref{eqn:state}) are not active in the time interval
$[t_{0},t_{f}]$. The Hamiltonian (\ref{eqn:H1}) then reduces to
\begin{equation}
\begin{aligned} H(\bm x,\bm\lambda, \bm u) = \mathcal{C}(\bm x, \bm u, t) + \bm\lambda^T(A\bm x + B\bm u) + \beta \end{aligned} \label{eqn:H2}%
\end{equation}
Observing that the terminal constraints (\ref{eqn:target}) $\bm\psi:=\bm
x-\bar{\bm X}=0$ are not explicit functions of time, the transversality
condition \citet{Bryson1969} is
\begin{equation}
\left.  H(\bm x(t),\bm\lambda(t),\bm u(t))\right\vert _{t=t_{f}}=0
\label{Transversality}%
\end{equation}
with $\bm\lambda(t_{f})=[(\bm\nu^{T}\frac{\partial\bm\psi}{\partial\bm x}%
)^{T}]_{t=t_{f}}$ as the costate boundary condition, where $\bm\nu$ denotes a
vector of Lagrange multipliers. The Euler-Lagrange equations become:
\begin{equation}
\dot{\bm\lambda}=-\frac{\partial H}{\partial\bm x}=-\frac{\partial
C(\bm x,\bm u,t)}{\partial\bm x}-A^{T}\bm\lambda, \label{EulerEqX}%
\end{equation}
and the necessary condition for optimality is
\begin{equation}
\frac{\partial H}{\partial\bm u}=\frac{\partial C(\bm x,\bm u,t)}%
{\partial\bm u}+B^{T}\bm\lambda=0. \label{OptimalityCondition}%
\end{equation}

With (\ref{eqn:H2})-(\ref{OptimalityCondition}), the initial state of system
(\ref{eqn:affine_noi}), and the terminal constraint $\bm x(t_{f})=\bar{\bm X}%
$, we can derive an unconstrained optimal state trajectory $\bm x^{\ast}(t)$
and optimal control $\bm u^{\ast}(t)$, $t\in\lbrack t_{0},t_{f}]$, for Problem
\ref{prob:general}.

When one or more constraints in (\ref{eqn:control}), (\ref{eqn:safetycons}),
(\ref{eqn:state}) become active in the time interval $[t_{0},t_{f}]$, we use
the interior point analysis \citet{Bryson1969} to determine the conditions that
must hold on a constrained arc entry point and exit point (if one exists prior
to $t_{f}$). We can then determine the optimal entry and exit points, as well
as the constrained optimal control $\bm u^{\ast}(t)$ and optimal state
trajectory $\bm x^{\ast}(t)$, $t\in\lbrack t_{0},t_{f}]$. Depending on the
computational complexity involved in deriving the complete constrained optimal
solution, we can specify a planned reference control $\bm u_{ref}(t)$ and
state trajectory $\bm x_{ref}(t)$, $t\in\lbrack t_{0},t_{f}]$. For example, we
may just plan for a safety-constrained solution and omit the state and control
constraints (\ref{eqn:control}), (\ref{eqn:state}), or even plan for only the
unconstrained optimal solution to simplify the trajectory planning process.

\subsection{Safety-Critical Optimal Control with HOCBFs}
\label{sec:scoc}

We now introduce a method that tracks the planned optimal control and state
trajectory while guaranteeing the satisfaction of all constraints
(\ref{eqn:control}), (\ref{eqn:safetycons}), (\ref{eqn:state}) in Problem
\ref{prob:general}.

As detailed in Sec. \ref{sec:oc}, we use $\bm u^{\ast}(t)$ and $\bm x^{\ast
}(t)$, $t\in\lbrack t_{0},t_{f}]$, to denote the optimal control and state
trajectory derived under no active constraints or with some (or all) of the
constraints active, depending on the associated computational complexity
considered acceptable in a particular setting. We can then reformulate
(\ref{eqn:gcost}) as the following optimization problem:%

\begin{equation}
\setlength{\abovedisplayskip}{1pt}\setlength{\belowdisplayskip}{1pt}\min
_{\bm u(t)}\int_{t_{0}}^{t_{f}}||\bm u(t)-\bm u_{ref}(t)||^{2}dt
\label{eqn:t2}%
\end{equation}
subject to (\ref{eqn:control}), (\ref{eqn:safetycons}), (\ref{eqn:state}),
where%
\begin{equation}
\bm u_{ref}(t)=F_{U}(\bm u^{\ast}(t),\bm x^{\ast}(t),\bm x(t))
\label{uref_general}%
\end{equation}
is a specific function of the optimal control and state trajectory, as well as
the actual state under noise $\bm w$ from (\ref{eqn:affine_noi}). A typical
choice for $F_{U}(\bm u^{\ast}(t),\bm x^{\ast}(t),\bm x(t))$ is
\begin{equation}
\bm u_{ref}(t)=e^{\sum_{j=1}^{n}\frac{x_{j}^{\ast}(t)-x_{j}(t)}{\sigma_{j}}%
}\bm u^{\ast}(t), \label{eqn:track}%
\end{equation}
where $x_{j}(t),$ $j\in\{1,\dots,n\}$ denote the observed state variables
under noise $\bm w$ from (\ref{eqn:affine_noi}), $x_{j}^{\ast}(t),$
$j\in\{1,\dots,n\},$ $u_{i}^{\ast}(t),$ $i\in\{1,\dots,q\}$ denote the optimal
state and control from the last subsection, and $\sigma_{j}>0,$ $j\in
\{1,\dots,n\}$ are weight parameters. In (\ref{eqn:track}), the sign of the
term $x_{j}^{\ast}(t)-x_{j}(t)$ depends on whether $x_{j}(t)$ is increasing
with $u_{i}(t)$. In particular, when $x_{j}(t)>x_{j}^{\ast}(t),$ for all
$j\in\{1,\dots,n\}$, we have $u_{i}(t)<u_{i}^{\ast}(t)$ and the state errors
can be automatically eliminated. If $x_{j}(t)<x_{j}^{\ast}(t),$ for all
$j\in\{1,\dots,n\}$, the state errors can similarly be automatically
eliminated. However, when $x_{j}(t)>x_{j}^{\ast}(t)$ and $x_{j+1}%
(t)<x_{j+1}^{\ast}(t)$, we may wish to enforce $u_{i}(t)<u_{i}^{\ast}(t),
i\in\{1,\dots, q\}$. Thus, it is desirable that $\sigma_{j}<\sigma_{j+1}$
(similarly, when $x_{j}(t)<x_{j}^{\ast}(t)$ and $x_{j+1}(t)>x_{j+1}^{\ast}%
(t)$). In summary, we select $\sigma_{j}>0,$ $j\in\{1,\dots,n\}$ such that
$\sigma_{j}<\sigma_{j+1},$ $j\in\{1,\dots,n-1\}$.

Alternative forms of (\ref{uref_general}) include
\begin{equation}
\bm u_{ref}(t)=\sum_{j\in\{1,\dots,n\}}\frac{x_{j}^{\ast}(t)}{x_{j}%
(t)}\bm u^{\ast}(t) \label{eqn:track2}%
\end{equation}
and the state feedback tracking control approach from \citet{Khalil2002}:
\begin{equation}
\bm u_{ref}(t)=\bm u^{\ast}(t)+\sum_{j=1}^{n}k_{j}(x_{j}^{\ast}(t)-x_{j}(t)),
\label{eqn:trackbk}%
\end{equation}
where $k_{j}>0,\in\{1,\dots,n\}$. Clearly, there are several possible choices
for the form of $\bm u_{ref}(t)$ which may depend on the specific application
of interest.

We emphasize that the cost (\ref{eqn:t2}) is subject to all the constraints
(\ref{eqn:control}), (\ref{eqn:safetycons}), (\ref{eqn:state}). We use HOCBFs
to implement these constraints, as well as CLFs to better track the optimal
state $\bm x^{\ast}(t)$, as shown in the following subsections.

\subsubsection{Optimal State Tracking}

\label{sec:statetrack} First, we aim to track the optimal state $\bm x^{\ast
}(t)$ obtained in Sec. \ref{sec:oc} using CLFs. We can always find a state
variable $x_{k},$ $k\in\{1,\dots,n\}$ in $\bm x$ that has relative degree one
(assume $x_{k}$ is the output) with respect to system (\ref{eqn:affine_noi}).
This is because we only take the Lie derivative of the Lyapunov function once
in the CLF constraint (\ref{eqn:clfp}). Then, we define a controller aiming to
drive $x_{k}(t)$ to $x_{ref}(t)$ where $x_{ref}(t)$ is of the form%
\begin{equation}
x_{ref}(t)=F_{X}(\bm x^{\ast}(t),\bm x(t)) \label{xref_general}%
\end{equation}
A typical choice analogous to (\ref{eqn:track}) is
\begin{equation}
x_{ref}(t)=e^{\sum_{j\in\{1,\dots,n\}\setminus k}\frac{x_{j}^{\ast}%
(t)-x_{j}(t)}{\sigma_{j}}}x_{k}^{\ast}(t) \label{StateTrackAno}%
\end{equation}
where $\sigma_{j}>0,$ $j\in\{1,\dots,n\}\setminus k$ {and
$\{1,\dots,n\}\setminus k$ denotes excluding $k$ from the set $\{1,\dots,n\}$%
}. An alternative form analogous to (\ref{eqn:track2}) is
\begin{equation}
x_{ref}(t)=\sum_{j\in\{1,\dots,n\}\setminus k}\frac{x_{j}^{\ast}(t)}{x_{j}%
(t)}x_{k}^{\ast}(t) \label{StateTrack}%
\end{equation}
where $x_{j}^{\ast}(t),$ $j\in\{1,\dots,n\}\setminus k$ are the (unconstrained
or constrained) optimal state trajectories from the Section
\ref{sec:optimalsolution}. In (\ref{StateTrack}), if $x_{j}(t)>x_{j}^{\ast
}(t)$, then $x_{ref}(t)<x_{k}^{\ast}(t)$, thus automatically reducing (or
eliminating) the tracking error. Note that while $x_{ref}(t)$ in
(\ref{StateTrack}) depends heavily on the exact value of $x_{j}(t)$, an
advantage of (\ref{StateTrackAno}) is that it allows $x_{ref}(t)$ to depend
only on the error. Clearly, we can define different tracking forms instead of
(\ref{StateTrack}) and (\ref{StateTrackAno}) depending on the specific
characteristics of an application.

Using a specific selected form of $x_{ref}(t)$, we can now proceed as in Def.
\ref{def:clf} and define an output $y_{k}(t):=x_{k}(t)-x_{ref}(t)$ for the
state variable $x_{k}$ which has relative degree one. Accordingly, we define a
CLF $V(y_{k}(t))=y_{k}^{2}(t)$ with $c_{1}=c_{2}=1,c_{3}=\epsilon>0$ as in
Def. \ref{def:clf}. Then, any control input $\bm u(t)$ should satisfy, for all
$t\in\lbrack t_{0},t_{f}]$,
\begin{equation}
\begin{aligned} L_fV(y_k(t)) + L_gV(y_k(t))\bm u(t) + \epsilon V(y_k(t)) \leq \delta_k(t) \end{aligned}\label{LyapunovTrack}%
\end{equation}
where $\delta_{k}(t)$ is a relaxation variable (to be minimized as explained
in the sequel) enabling the treatment of the requirement $x_{k}(t)=x_{ref}(t)$
as a soft constraint. Note that we may also identify other state variables
with relative degree one and define multiple CLFs to better track the optimal
state. {Note that (\ref{LyapunovTrack}) does not include any
(unknown) noise term. Also note that selecting a larger $\epsilon$ can improve
the state convergence rate
\citet{Aaron2012}.}

\subsubsection{Safety Constraints and State Limitations}

Next, we use HOCBFs to map the safety constraints (\ref{eqn:safetycons}) and
state limitations (\ref{eqn:state}) from the state $\bm x(t)$ to the control
input $\bm u(t)$. Let $b_{j}(\bm x),$ $j\in S_{o}$, be the HOCBF corresponding
to the $j$th safety constraint. In addition, let $b_{i,\max}(\bm x)=x_{i,\max
}-x_{i}$ and $b_{i,\min}(\bm x)=x_{i}-x_{i,\min},$ $i\in\{1,\dots,n\},$ be the
HOCBFs for all state limitations, where $\bm x_{\max}=(x_{1,\max}%
,\dots,x_{n,\max}),$ $\bm x_{\min}=(x_{1,\min},\dots,x_{n,\min})$. The
relative degrees of $b_{i,\max}(\bm x),$ $b_{i,\min}(\bm x),$ $i\in
\{1,\dots,n\}$ are $m_{i}$, and the relative degrees of $b_{j}(\bm x),$ $j\in
S_{o}$ are $m_{j}$. Therefore, in Definition \ref{def:hocbf}, we choose HOCBFs
with $m=m_{i}$ or $m_{j}$, including the penalty factors $p_{i,\min}>0,
p_{i,\max}>0, p_{i,safe}>0$ (see discussion after Definition \ref{def:hocbf})
for all the class $\mathcal{K}$ functions. Following (\ref{eqn:constraint}),
any control input $u_{i}(t)$ should satisfy {\small
\begin{equation}
L_{f}^{m_{j}}b_{j}(\bm x)\!+\!L_{g}L_{f}^{m_{j}-1}b_{j}%
(\bm x)\bm u\!+\!S(b_{j}(\bm x))\!+\!p_{i,safe}\alpha_{m_{j}}(\psi_{m_{j}%
-1}(\bm x))\geq0,j\in S_{o}, \label{eqn:safety0}%
\end{equation}
}{\small
\begin{subequations}\label{eqn:safety1}
\begin{align}
L_{f}^{m_{i}}b_{i,\max}(\bm x) \!+\! L_{g}L_{f}^{m_{i}-1}b_{i,\max
}(\bm x)\bm u \!+\! S(b_{i,\max}(\bm x)) \!+\! p_{i,\max}\alpha_{m_{i}}%
(\psi_{m_{i}-1}(\bm x)) \geq0,\\
L_{f}^{m_{i}}b_{i,\min}(\bm x) \!+\! L_{g}L_{f}^{m_{i}-1}b_{i,\min
}(\bm x)\bm u \!+\! S(b_{i,\min}(\bm x)) \!+\! p_{i,\min}\alpha_{m_{i}}%
(\psi_{m_{i}-1}(\bm x)) \geq0,
\end{align}
\end{subequations}
}for all $t\in\lbrack t_{0},t_{f}],$ $i\in\{1,\dots,n\}$. Note that $\bm u\in
U$ in (\ref{eqn:control}) are already constraints on the control inputs,
hence, we do not need to use HOCBFs for them.

\subsubsection{Joint Optimal and HOCBF (OCBF) Controller}

Using the HOCBFs and CLFs introduced in the last two subsections, we can
reformulate objective (\ref{eqn:t2}) in the form: {\small
\begin{equation}
\int_{t_{0}}^{t_{f}}\!\left(  \beta\delta_{k}^{2}(t)\!+||\bm u(t)-\bm u_{ref}%
(t)||^{2}\right)  dt, \label{eqn:objT}%
\end{equation}
}subject to (\ref{eqn:affine_noi}), (\ref{LyapunovTrack}), (\ref{eqn:safety0}%
), (\ref{eqn:safety1}), and (\ref{eqn:control}), the initial conditions
$\bm x(t_{0})$, and given $t_{0}$. Thus, we have combined the HOCBF method and
the optimal control solution by using (\ref{uref_general}) to link the optimal
state and control to $\bm u_{ref}(t)$, and using (\ref{xref_general}) in the
CLF $(x(t)-x_{ref}(t))^{2}$ to combine with (\ref{eqn:t}). We refer to the
resulting control $\bm u(t)$ in (\ref{eqn:objT}) as the \emph{OCBF control}.

Finally, we partition the continuous time interval $[t_{0},t_{f}]$ into equal
time intervals $\{[t_{0}+\omega\Delta t,t_{0}+(\omega+1)\Delta t)\},$
$\omega=0,1,2,\dots$. In each interval $[t_{0}+\omega\Delta t,t_{0}%
+(\omega+1)\Delta t)$, we assume the control is constant and find a solution
to the optimization problem in (\ref{eqn:objT}) using the CLF $y_{k}%
=(x_{k}(t)-x_{ref}(t))^{2}$ and associated relaxation variable $\delta_{k}%
(t)$. Specifically, at $t=t_{0}+\omega\Delta t$ ($\omega=0,1,2,\dots$), we
solve \
\begin{equation}
\underset{t=t_{0}+\omega\Delta t}{\mathbf{QP:}}(\bm u^{\star}(t),\delta
_{k}^{\star}(t))=\mathop{\arg\min}_{\bm u(t),\delta_{k}(t)}\left[  \beta
\delta_{k}^{2}(t)\!+||\bm u(t)-\bm u_{ref}(t)||^{2}\right]  \label{QP}%
\end{equation}
subject to
\begin{equation}
A_{\text{clf}}[\bm u(t),\delta_{k}(t)]^{T}\leq b_{\text{clf}} \label{QPcontr1}%
\end{equation}%
\begin{equation}
A_{\text{cbf\_lim}}[\bm u(t),\delta_{k}(t)]^{T}\leq b_{\text{cbf\_lim}}
\label{QPcontr2}%
\end{equation}%
\begin{equation}
A_{\text{cbf\_safe}}[\bm u(t),\delta_{k}(t)]^{T}\leq b_{\text{cbf\_safe}}
\label{QPcontr3}%
\end{equation}
The constraint parameters $A_{\text{clf}},$ $b_{\text{clf}}$ pertain to the
reference state tracking CLF constraint (\ref{LyapunovTrack}):
\begin{equation}
\begin{aligned} A_{\text{clf}} &= [L_gV(y_k(t)),\qquad -1],\\ b_{\text{clf}} &= -L_fV(y_k(t)) - \epsilon V(y_k(t)). \end{aligned}
\end{equation}
On the other hand, the constraint parameters $A_{\text{cbf\_lim}},$
$b_{\text{cbf\_lim}}$ capture the state HOCBF constraints (\ref{eqn:safety1})
and the control bounds (\ref{eqn:control}):
\begin{equation}
\begin{aligned} A_{\text{cbf\_lim}} &= \left[\begin{array}{cc} -L_gL_f^{m_i-1}b_{i,\max}(\bm x(t)), & 0\\ -L_gL_f^{m_i-1}b_{i,\min}(\bm x(t)), & 0\\ 1, & 0\\ -1, & 0 \end{array} \right],\\ b_{\text{cbf\_lim}} &= \left[\begin{array}{c} L_f^{m_i}b_{i,\max}(\bm x) + S(b_{i,\max}(\bm x)) + p_{i,\max}\alpha_{m_i}(\psi_{m_i-1}(\bm x))\\ L_f^{m_i}b_{i,\min}(\bm x) + S(b_{i,\min}(\bm x)) + p_{i,\min}\alpha_{m_i}(\psi_{m_i-1}(\bm x))\\ \bm u_{max}\\ -\bm u_{min} \end{array} \right]. \end{aligned}
\end{equation}
for all $i\in\{1,\dots,n\}$. Finally, the constraint parameters
$A_{\text{cbf\_safe}},$ $b_{\text{cbf\_safe}}$ capture the safety HOCBF
constraints (\ref{eqn:safety0}), for all $j\in S_{o}$:
\begin{equation}
\begin{aligned} A_{\text{cbf\_safe}}\! &=\! \left[\begin{array}{cc} -L_gL_f^{m_j-1}b_j(\bm x), & 0 \end{array} \right],\\ b_{\text{cbf\_safe}}\! &=\!\begin{array}{c} L_f^{m_j}b_j(\bm x) + S(b_j(\bm x)) + p_{j,safe}\alpha_{m_j}(\psi_{m_j-1}(\bm x)). \end{array} \end{aligned}
\end{equation}
From a computational complexity point of view, it normally takes a fraction of
a second (see explicit results in Sec. \ref{sec:simulation}) to solve
(\ref{QP}) in MATLAB, rendering the OCBF controller very efficient for
real-time implementation. After solving each (\ref{QP}) we obtain an optimal
OCBF control $\bm u^{\star}(t)$, not to be confused with a solution of the
original optimal control problem (\ref{eqn:gcost}). We then update
(\ref{eqn:affine_noi}) and apply it to all $t\in\lbrack t_{0}+\omega\Delta
t,t_{0}+(\omega+1)\Delta t)$.

\begin{remark}
If we can find conditions such that the constraints are not active, then we
can simply track the unconstrained optimal control and state. This simplifies
the implementation of the optimal trajectory planning without considering
constraints, i.e., we can directly apply $\bm u_{ref}$ in (\ref{uref_general})
as the control input of system (\ref{eqn:affine_noi}) instead of solving
(\ref{QP}). {The feasibility of QP (\ref{QP}) can be improved
through smaller $p_{i,\min},p_{i,\max},p_{j,safe}$ at the expense of possibly
shrinking the initial feasible set \citet{Xiao2019}.}
\end{remark}

\subsection{Constraint Violation Due to Noise}

The presence of noise in the dynamics (\ref{eqn:affine_noi}) will generally
result in the violation of the constraints (\ref{eqn:state}) or
(\ref{eqn:safetycons}), which prevents the HOCBF method from satisfying the
forward invariance property \citet{Xiao2019}. Therefore, we seek to minimize
the time during which such a constraint is violated.

\subsubsection{Relative Degree One Constraints}

Suppose that a constraint $b(\bm x(t))\geq0$ (one of the constraints in
(\ref{eqn:state}),(\ref{eqn:safetycons})) has relative degree one for system
(\ref{eqn:affine_noi}). Let us first assume that $\bm w$ in
(\ref{eqn:affine_noi}) is bounded by $|\bm w|\leq\bm W$, where $W>\bm0$
(componentwise). Then, the following modified CBF constraint
\citet{Lindemann2019} can guarantee that $b(\bm x(t))\geq0$ is always satisfied
under $|\bm w|\leq\bm W$:
\begin{equation}
L_{f}b(\bm x(t))+L_{g}b(\bm x(t))\bm u(t)+\alpha(b(\bm x(t)))-\left\vert
\frac{d b(\bm x(t))}{d\bm x}\right\vert \bm W\geq0. \label{eqn:cbfnoise}%
\end{equation}
The HOCBF constraint (\ref{eqn:constraint}) with $m=1$ is equivalent to
$L_{f}b(\bm x(t))+L_{g}b(\bm x(t))\bm u(t)+\alpha(b(\bm x(t)))+\frac{d
b(\bm x(t))}{d\bm x}\bm w\geq0$ if we take the derivative of $b(\bm
x(t))$ along the noisy dynamics (\ref{eqn:affine_noi}). Thus, the satisfaction
of (\ref{eqn:cbfnoise}) implies the satisfaction of this constraint. Note that
the modified CBF constraint (\ref{eqn:cbfnoise}) is conservative since it
always considers the (deterministic) noise bound $\bm W$.

Next, suppose a bound $\bm W$ is unknown, in which case we can proceed as
follows. Assume the constraint is violated at time $t_{1}\in\lbrack
t_{0},t_{f}]$ due to noise, i.e., we have $b(\bm x(t_{1}))<0$. We need to
ensure that $b(\bm x(t))$ is strictly increasing after time $t_{1}$, i.e.,
$\dot{b}(\bm x(t))\geq c(t)$, where $c(t)$ is positive and is desired to take
the largest possible value maintaining the feasibility of the QP (\ref{QP}),
i.e., we wish to maximize $c(t)$ at each time step (alternatively, we can set
$c(t)=c>0$ as a positive constant). Using Lie derivatives, we evaluate the
change in $b(\bm x(t))$ along the flow defined by the system state vector.
Then, any control $\bm u(t)$ must satisfy
\begin{equation}
L_{f}b(\bm x(t))+L_{g}b(\bm x(t))\bm u(t)\geq c(t) \label{eqn:Lie}%
\end{equation}
since we wish to maximize $c(t)$ so that $b(\bm x(t))$ is strictly increasing
even if the system is subject to the worst possible noise case. For this
reason, in what follows we assume that the random process $\bm w(t)$ in
(\ref{eqn:affine_noi}) is characterized by a probability density function with
finite support and we incorporate the maximization of $c(t)$ into the cost
(\ref{eqn:objT}) as follows: {\small
\begin{equation}
\min_{\bm u(t),\delta_{k}(t),c(t)}\int_{t_{0}}^{t_{f}}\!\left(  \beta
\delta_{k}^{2}(t)\!+\!||\bm u-\bm u_{ref}||^{2}-Kc(t)\right)  dt,
\label{eqn:energyvio}%
\end{equation}
} where $K>0$ is a large scalar weight parameter.

Note that several constraints may be violated at the same time. Starting from
$t_{1}$, we apply the constraint (\ref{eqn:Lie}) to the HOCBF optimizer
instead of the HOCBF constraint (\ref{eqn:constraint}), and $b(\bm x(t))$ will
be positive again in finite time since it is strictly increasing. When
$b(\bm
x(t))$ becomes positive again at $t_{2}\in[t_{1},t_{f}]$, we can once again
apply the HOCBF constraint (\ref{eqn:constraint}).

\subsubsection{High Relative Degree Constraints}

If a constraint $b(\bm x(t))\geq0$ is such that $b:\mathbb{R}^{n}%
\rightarrow\mathbb{R}$ has relative degree $m>1$ for (\ref{eqn:affine_noi}),
we can no longer find a modified CBF constraint as in (\ref{eqn:cbfnoise})
that guarantees $b(\bm x(t))\geq0$ under noise $\bm w$. This is because we
need to know the bounds of the derivatives of $\bm w$ as $b(\bm x(t))$ will be
differentiated $m$ times. In other words, we need to recursively drive
$b^{(i)}(\bm x(t))=\frac{d^{i}b(\bm x(t))}{dt^{i}}$ to be positive from $i=m$
to $i=1$ after it is violated at some time $t\in\lbrack t_{0},t_{f}]$.
Therefore, we need knowledge of the positive degree of $b(\bm x(t))$ at $t$
which is defined as follows.

\begin{definition}
\label{def:positive} (\textit{Positive degree}) The positive degree $\rho(t)$
of a relative degree $m$ function $b:\mathbb{R}^{n}\rightarrow\mathbb{R}$ at
time $t$ is defined as:
\end{definition}
\begin{equation}
\rho(t)\!:=\!\left\{
\begin{array}
[c]{lll}%
\mathop{\min}\limits_{i\in\{0,\ldots,m\!-\!1\}:b^{(i)}(\bm x(t))>0} i, &
\text{if }\exists i\in\{0,\ldots,m\!-\!1\} & \\
m & otherwise &
\end{array}
\right.  \label{eqn:positive}%
\end{equation}
If $b^{(i)}(\bm x(t))\leq0,$ \ for all $i\in\{0,\dots,m-1\}$, $\bm u(t)$
shows up in $b^{(m)}(\bm x(t))$ since the function $b$ has relative degree $m$
for system (\ref{eqn:affine_noi}). Therefore, we may choose a proper control
input $\bm u(t)$ such that $b^{(m)}(\bm x(t))>0$, and, in this case,
$\rho(t)=m$. The positive degree of $b(\bm x(t))$ at time $t$ is 0 if
$b(\bm x(t))>0$.

Letting $\psi_{0}(\bm x,t):=b(\bm x(t))$, we can construct a sequence of
functions $\psi_{i}:\mathbb{R}^{n}\rightarrow\mathbb{R},\forall i\in
\{1,\dots,m\}$ similar to (\ref{eqn:functions}):
\begin{equation}
\begin{aligned} \psi_i(\bm x) :=& \left\{ \begin{array}{lll} \dot\psi_{i-1}, \text{ if }i<\rho(t), \\ \dot\psi_{i-1}(\bm x) - \varepsilon, \text{ if }i=\rho(t), \\ \dot\psi_{i-1}(\bm x) + \alpha_{i}(\psi_{i-1}(\bm x)), \text{ otherwise}. \end{array} \right.\\ \end{aligned} \label{eqn:gfunctions}%
\end{equation}
where $\alpha_{i}(\cdot),i\in\{1,\dots,m\}$, denote class $\mathcal{K}$
functions of their argument and $\varepsilon>0$ is a constant. {We
may choose $\varepsilon\geq\left|  \frac{d\psi_{i-1}(\bm x)}{d\bm x}\right|
\bm W$ if $\bm w$ is bounded as in (\ref{eqn:cbfnoise}).}

We can then define a sequence of sets $C_{i}$ similar to (\ref{eqn:sets})
associated with the $\psi_{i-1}(\bm x),i\in\{1,\dots,m\}$ functions in
(\ref{eqn:gfunctions}). We replace the definitions of $\psi_{i-1}(\bm
x),C_{i},$ $i\in\{1,\dots,m\}$ in Def. \ref{def:hocbf} to define $b(\bm x)$ to
be a HOCBF.

If $\rho(t)=m$, then $\psi_{m}(\bm x(t))=\dot{\psi}_{m-1}(\bm
x(t))-\varepsilon\geq0$, which is equivalent to the HOCBF constraint
(\ref{eqn:constraint}). The control $\bm u$ that satisfies $\dot{\psi}%
_{m-1}(\bm x(t))\geq\varepsilon>0$ will drive $\psi_{m-1}(\bm x(t))>0$ in
finite time. Otherwise, since $\psi_{\rho(t)}(\bm x(t))>0$ according to Def.
\ref{def:positive}, we can always choose proper class $\mathcal{K}$ functions
$\alpha_{i}(\cdot),i\in\{\rho(t)+1,\dots,m\}$ such that $\psi_{i}(\bm x)\geq
0$, i.e., we can construct a non-empty set $C_{\rho(t)+1}\cap\dots\cap C_{m}$
\citet{Xiao2019}. By Theorem \ref{thm:hocbf}, the set $C_{\rho(t)+1}\cap
\dots\cap C_{m}$ is forward invariant if the HOCBF constraint
(\ref{eqn:constraint}) is satisfied. In other words, $\psi_{\rho(t)}(\bm
x(t))\geq0$ is guaranteed. Since $\psi_{\rho(t)}(\bm x(t))=\dot{\psi}%
_{\rho(t)-1}(\bm x(t))-\varepsilon$, then $\dot{\psi}_{\rho(t)-1}(\bm
x(t))\geq\varepsilon>0$. The function $\psi_{\rho(t)-1}(\bm x(t))$ will become
positive in finite time, and the positive degree of $b(\bm x(t))$ will
decrease by one. Proceeding recursively at most $m$ times, eventually the
positive degree of $b(\bm x(t))$ will be 0, i.e., the original constraint
$b(\bm x(t))>0$ is satisfied in finite time. The time needed for the
constraint $b(\bm x(t))>0$ to be satisfied depends on the magnitude of
$\varepsilon$.

\section{TRAFFIC MERGING PROBLEM}

\label{sec:prob}

In the rest of the paper, we apply the OCBF framework developed thus far to
the traffic merging problem where the goal is to optimally control CAVs
approaching a merging point while guaranteeing safety constraints at all times.

\begin{figure}[ptbh]
\centering
\includegraphics[scale=0.22]{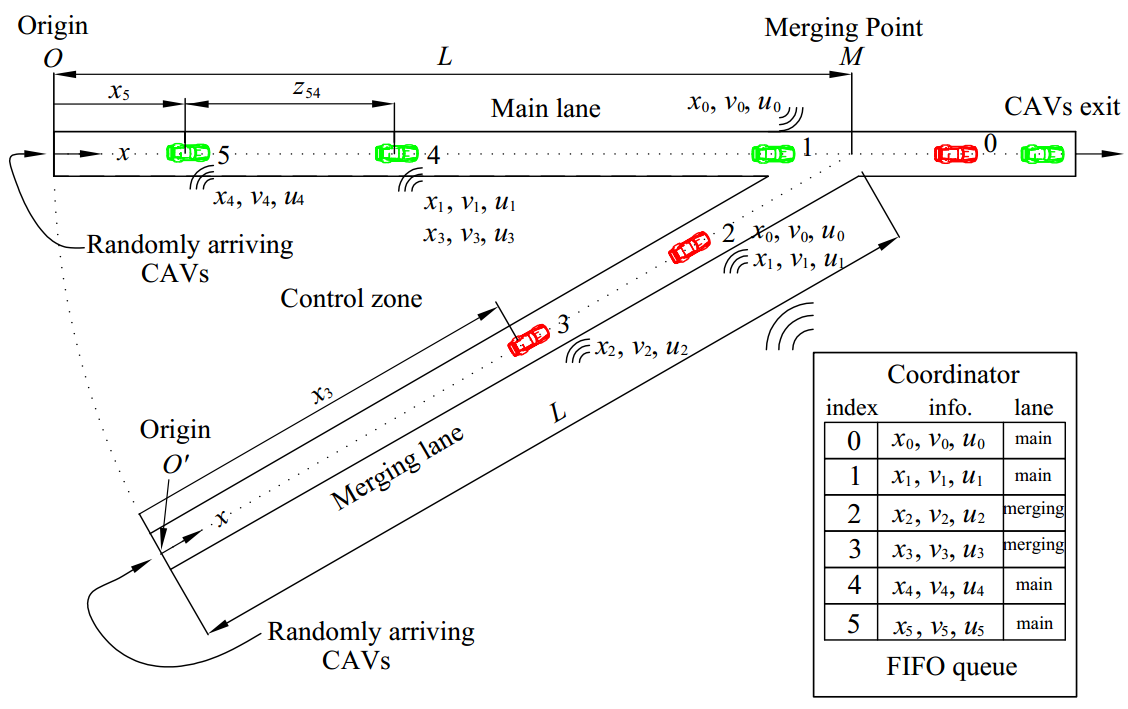} \caption{The merging problem}%
\label{fig:merging}%
\end{figure}

The merging problem arises when traffic must be joined from two different
roads, usually associated with a main lane and a merging lane as shown in
Fig.\ref{fig:merging}. We consider the case where all traffic consists of CAVs
randomly arriving at the two lanes joined at the Merging Point (MP) $M$ where
a collison may occur. The segment from the origin $O$ or $O^{\prime}$ to the
MP $M$ has a length $L$ for both lanes, and is called the Control Zone (CZ).
We assume that CAVs do not overtake each other in the CZ. A coordinator is
associated with the MP whose function is to maintain a First-In-First-Out
(FIFO) queue of CAVs based on their arrival time at the CZ and enable
real-time communication with the CAVs that are in the CZ as well as the last
one leaving the CZ. The FIFO assumption imposed so that CAVs cross the MP in
their order of arrival is made for simplicity and often to ensure fairness,
but can be relaxed through dynamic resequencing schemes, e.g., as described in
\citet{Wei2020ACC}. Let $S(t)$ be the set of FIFO-ordered indices of all CAVs
located in the CZ at time $t$ along with the CAV (whose index is 0 as shown in
Fig. \ref{fig:merging}) that has just left the CZ. Let $N(t)$ be the
cardinality of $S(t)$. Thus, if a CAV arrives at time $t$ it is assigned the
index $N(t)$. All CAV indices in $S(t)$ decrease by one when a CAV passes over
the MP and the vehicle whose index is $-1$ is dropped.

We review next the optimal merging control problem as presented in
\citet{Wei2019ACC} so as to apply the OCBF framework to it. The vehicle
dynamics for each CAV $i\in S(t)$ along the lane to which it belongs take the
form
\begin{equation}
\left[
\begin{array}
[c]{c}%
\dot{x}_{i}(t)\\
\dot{v}_{i}(t)
\end{array}
\right]  =\left[
\begin{array}
[c]{c}%
v_{i}(t)\\
u_{i}(t)
\end{array}
\right]  \label{VehicleDynamics}%
\end{equation}
where $x_{i}(t)$ denotes the distance to the origin $O$ ($O^{\prime}$) along
the main (merging) lane if the vehicle $i$ is located in the main (merging)
lane, $v_{i}(t)$ denotes the velocity, and $u_{i}(t)$ denotes the control
input (acceleration). We consider two objectives for each CAV subject to three
constraints, as detailed next.

\textbf{Objective 1} (Minimizing travel time): Let $t_{i}^{0}$ and $t_{i}^{M}$
denote the time that CAV $i\in S(t)$ arrives at the origin $O$ or $O^{\prime}$
and the MP $M$, respectively. We wish to minimize the travel time $t_{i}%
^{M}-t_{i}^{0}$ for CAV $i$.

\textbf{Objective 2} (Minimizing energy consumption): We also wish to minimize
energy consumption for each CAV $i\in S(t)$ expressed as
\begin{equation}
J_{i}(u_{i}(t))=\int_{t_{i}^{0}}^{t_{i}^{M}}C(u_{i}(t))dt,
\end{equation}
where $C(\cdot)$ is a strictly increasing function of its argument.

\textbf{Constraint 1} (Safety constraints): Let $i_{p}$ denote the index of
the CAV which physically immediately precedes $i$ in the CZ (if one is
present). We require that the distance $z_{i,i_{p}}(t):=x_{i_{p}}(t)-x_{i}(t)$
be constrained by the speed $v_{i}(t)$ of CAV $i\in S(t)$ so that
\begin{equation}
z_{i,i_{p}}(t)\geq\varphi v_{i}(t)+\delta_{0},\text{ \ }\forall t\in\lbrack
t_{i}^{0},t_{i}^{M}], \label{Safety}%
\end{equation}
where $\varphi$ denotes the reaction time (as a rule, $\varphi=1.8$ is used,
e.g., \citet{Vogel2003}). If we define $z_{i,i_{p}}$ to be the distance from
the center of CAV $i$ to the center of CAV $i_{p}$, then $\delta_{0}$ is a
constant determined by the length of these two CAVs (generally dependent on
$i$ and $i_{p}$ but taken to be a constant over all CAVs for simplicity).

\textbf{Constraint 2} (Safe merging): There should be enough safe space at the
MP $M$ for a merging CAV to cut in, i.e.,
\begin{equation}
z_{1,0}(t_{1}^{M})\geq\varphi v_{1}(t_{1}^{M})+\delta_{0}. \label{SafeMerging}%
\end{equation}

\textbf{Constraint 3} (Vehicle limitations): Finally, there are constraints on
the speed and acceleration for each $i\in S(t)$, i.e.,
\begin{equation}
\begin{aligned} v_{\min} \leq v_i(t)\leq v_{\max}, \forall t\in[t_i^0,t_i^M],\\ u_{\min}\leq u_i(t)\leq u_{\max}, \forall t\in[t_i^0,t_i^M], \end{aligned} \label{VehicleConstraints}%
\end{equation}
where $v_{\max}>0$ and $v_{\min}\geq0$ denote the maximum and minimum speed
allowed in the CZ, while $u_{\min}<0$ and $u_{\max}>0$ denote the minimum and
maximum control input, respectively.

The common way to minimize energy consumption is by minimizing the control
input effort $u_{i}^{2}(t)$. By normalizing travel time and $u_{i}^{2}(t)$,
and using $\alpha\in\lbrack0,1]$, we construct a convex combination as in
(\ref{eqn:cost}): {\small
\begin{equation}
\setlength{\abovedisplayskip}{2pt}\setlength{\belowdisplayskip}{2pt}\begin{aligned} J_i(u_i(t))= \int_{t_i^0}^{t_i^M}\left(\alpha + \frac{(1-\alpha)\frac{1}{2}u_i^2(t)}{\frac{1}{2}\max \{u_{\max}^2, u_{\min}^2\}}\right)dt \end{aligned}.\label{eqn:energyobja}%
\end{equation}
}If $\alpha=1$, then we solve (\ref{eqn:energyobja}) as a minimum time
problem. Otherwise, by defining $\beta:=\frac{\alpha\max\{u_{\max}^{2}%
,u_{\min}^{2}\}}{2(1-\alpha)}$ and multiplying the last equation by
$\frac{\beta}{\alpha}$, we have: {\small
\begin{equation}
\setlength{\abovedisplayskip}{1pt}\setlength{\belowdisplayskip}{1pt}J_{i}%
(u_{i}(t)):=\beta(t_{i}^{M}-t_{i}^{0})+\int_{t_{i}^{0}}^{t_{i}^{M}}\frac{1}%
{2}u_{i}^{2}(t)dt,\label{eqn:energyobj}%
\end{equation}
}where $\beta\geq0$ is a weight factor that can be adjusted to penalize travel
time relative to the energy cost. {Note that all the constraints in
the merging problem are with relative degree one.}

Similar to (\ref{eqn:affine_noi}), we will also include the possibility of
system model uncertainties, errors due to signal transmission, as well as
computation errors by adding two noise terms in (\ref{VehicleDynamics}) to
get
\begin{equation}
\left[
\begin{array}
[c]{c}%
\dot{x}_{i}(t)\\
\dot{v}_{i}(t)
\end{array}
\right]  =\left[
\begin{array}
[c]{c}%
v_{i}(t)+w_{i,1}(t)\\
u_{i}(t)+w_{i,2}(t)
\end{array}
\right]  \label{VehicleDynamicsNoise}%
\end{equation}
where $w_{i,1}(t),w_{i,2}(t)$ denote two random processes defined in an
appropriate probability space. 

\section{MERGING PROBLEM ANALYSIS}

\label{sec:ocbf}

In this section, we first review the decentralized optimal control (OC)
solution derived in \citet{Wei2019ACC} for those CAVs whose constraints in
(\ref{Safety})-(\ref{VehicleConstraints}) will not become active in the CZ.
This is to ensure that these solutions are indeed computationally efficient.
When one or more constraints becomes active, we use the CBF method to account
for these constraints and take the unconstrained optimal solution as
reference. When more complex objective functions, nonlinear dynamics, and
comfort are involved, we adapt the CBF method to such problems. In addition,
we show how we can deal with the constraint violation problem due to
perturbations, such as the noise in (\ref{VehicleDynamicsNoise}) and other
unknown random events.

We need to distinguish between the following two cases: $(i)$ $i_{p}=i-1$,
\textit{i.e.,} $i_{p}$ is the CAV immediately preceding $i$ in the FIFO queue
(such as CAV 3 or 5 in Fig. \ref{fig:merging}), and $(ii)$ $i_{p}<i-1$ (such
as CAV 2 or 4 in Fig. \ref{fig:merging}), which implies CAV $i-1$ is in a
different lane from $i$. We can solve the merging problem for all $i\in S(t)$
in a decentralized way, in the sense that CAV $i$ can solve it using only its
own local information (position, velocity and acceleration) along with that of
its \textquotedblleft neighbor\textquotedblright\ CAVs $i-1$ and $i_{p}$.
Observe that if $i_{p}=i-1$, then (\ref{SafeMerging}) is a redundant
constraint. Otherwise, we need to consider (\ref{Safety}) and
(\ref{SafeMerging}) independently.

Let $\bm x_{i}(t):=(x_{i}(t),v_{i}(t))$ be the state vector and $\bm\lambda
_{i}(t):=(\lambda_{i}^{x}(t),\lambda_{i}^{v}(t))$ be the costate vector (for
simplicity, in the sequel we omit explicit time dependence when no ambiguity
arises). The Hamiltonian for the merging problem with the state, control, and
safety constraints adjoined is
\begin{equation}
\begin{aligned} H_i(\bm x_i,\bm\lambda_i, u_i) = &\beta + \frac{1}{2}u_i^2\! +\! \lambda_i^xv_i+ \lambda_i^vu_i + \mu_i^a(u_i\! -\! u_{max}) \\&+ \mu_i^b(u_{min} - u_i) + \mu_i^c(v_i - v_{max}) \\&+ \mu_i^d(v_{min} - v_i) + \mu_i^e(x_i + \varphi v_i + \delta_0 -x_{i_p}) \end{aligned}
\end{equation}
The Lagrange multipliers $\mu_{i}^{a},\mu_{i}^{b},\mu_{i}^{c},\mu_{i}^{d}%
,\mu_{i}^{e}$ are positive when the constraints are active and become 0 when
the constraints are strict. Note that when the safety constraint
(\ref{Safety}) becomes active, the expression above involves $x_{i_{p}}(t)$ in
the last term. When $i=1$, the optimal trajectory is obtained without this
term, since (\ref{Safety}) is inactive over all $[t_{1}^{0},t_{1}^{M}]$. Thus,
once the solution for $i=1$ is obtained, $x_{1}^{\ast}$ is a given function of
time and available to $i=2$. Based on this information, the optimal trajectory
of $i=2$ is obtained. Similarly, all subsequent optimal trajectories for $i>2$
can be recursively obtained based on $x_{i_{p}}^{\ast}(t)$.

\subsection{CAVs with Unconstrained Optimal Control}

Assuming that (\ref{Safety}) and (\ref{VehicleConstraints}) remain inactive
over $[t_{i}^{0},t_{i}^{M}]$, and the safe merging constraint
(\ref{SafeMerging}) is not violated at $t_{i}^{M}$, we can obtain the
unconstrained optimal solution as shown in \citet{Wei2019ACC}:
\begin{equation}
u_{i}^{\ast}(t)=a_{i}t+b_{i} \label{Optimal_u}%
\end{equation}%
\begin{equation}
v_{i}^{\ast}(t)=\frac{1}{2}a_{i}t^{2}+b_{i}t+c_{i} \label{Optimal_v}%
\end{equation}%
\begin{equation}
x_{i}^{\ast}(t)=\frac{1}{6}a_{i}t^{3}+\frac{1}{2}b_{i}t^{2}+c_{i}t+d_{i}
\label{Optimal_x}%
\end{equation}
where $a_{i}$, $b_{i}$, $c_{i}$ and $d_{i}$ are integration constants obtained
by solving the following five nonlinear algebraic equations:
\begin{equation}
\begin{aligned} &\frac{1}{2}a_i\cdot(t_i^0)^2 + b_it_i^0 + c_i = v_i^0,\\ &\frac{1}{6}a_i\cdot(t_i^0)^3 + \frac{1}{2}b_i\cdot(t_i^0)^2 + c_it_i^0+d_i = 0,\\ &\frac{1}{6}a_i\cdot(t_i^M)^3 + \frac{1}{2}b_i\cdot(t_i^M)^2 + c_it_i^M+d_i = L,\\ &a_it_i^M + b_i = 0,\\ &\beta + \frac{1}{2}a_i^2\cdot(t_i^M)^2 + a_ib_it_i^M + a_ic_i = 0. \end{aligned} \label{OptimalSolInA}%
\end{equation}

Since we aim for the solution to the optimal merging problem to be obtained
on-board each CAV, it is essential that the computational cost of solving
these five algebraic equations for the integration constants in
(\ref{Optimal_u})-(\ref{Optimal_x}) be minimal. If MATLAB\ is used, it takes
less than 1 second to solve these equations (Intel(R) Core(TM) i7-8700 CPU @
3.2GHz 3.2GHz). On the other hand, when the constraints (\ref{Safety}),
(\ref{SafeMerging}), (\ref{VehicleConstraints}) become active, a complete OC
solution can still be obtained \citet{Wei2019ACC}, \citet{Malikopoulos2018}, but the
computation time varies between 3 and 30 seconds depending on whether $i_{p}$
is also safety-constrained or not. This motivates the derivation of conditions
such that these constraints do not become active in the CZ.

The following assumption requires that if two CAVs arrive too close to each
other, then the first one maintains its optimal terminal speed past the MP
until the second one crosses it as well. This is to ensure that the first
vehicle does not suddenly decelerate and cause the safety constraint to be
violated during the last segment of its optimal trajectory.

\begin{assump}
\label{asp:constant} For a given constant $\zeta= \frac{v_{i}^{*}(t_{i}^{M}%
)}{v_{i-1}^{*}(t_{i-1}^{M})}\varphi+ \frac{\delta_{0}}{v_{i-1}^{*}(t_{i-1}%
^{M})}$, any CAV $i-1\in S(t)$ such that $t_{i}^{M}-t_{i-1}^{M}<\zeta$
maintains a constant speed $v_{i-1}(t)=v_{i-1}^{\ast}(t_{i-1}^{M})$ for all
$t\in\lbrack t_{i-1} ^{M},t_{i}^{M}]$.
\end{assump}

Based on this mild assumption, the following theorems from \citet{Wei2019CDC}
ensure that the constraints (\ref{Safety}), (\ref{SafeMerging}),
(\ref{VehicleConstraints}) are satisfied. The first identfies simple to check
conditions such that the safety constraint (\ref{Safety}) will not become
active within the CZ and the second identifies conditions such that the safe
merging constraint (\ref{SafeMerging}) will not be violated at $t_{i}^{M}$.

\begin{theorem}
\label{thm:safety_general} \citet{Wei2019CDC} Under Assumption
\ref{asp:constant}, if $\exists\varepsilon\in(0,1]$ such that $\varepsilon
v_{i}^{0}\leq v_{i_{p}}^{0}$ and $t_{i}^{0}-t_{i_{p}}^{0}\geq\frac{\varphi
}{\varepsilon}+\frac{\delta_{0}}{\varepsilon v_{i}^{0}}+\frac{3L(1-\varepsilon
)}{v_{i_{p}}^{0}+2v_{i_{p}}^{*}(t_{i_{p}}^{M})}$, then, under optimal control
(\ref{Optimal_u}) for both $i$ and $i_{p}$, $z_{i,i_{p}}(t_{i}^{M})\geq\varphi
v_{i}(t_{i}^{M})+\delta_{0}$. Moreover, if $\exists t_{p}\in\lbrack t_{i}
^{0},t_{i_{p}}^{M})$ solved by $v_{i}(t_{p})+\varphi u_{i}(t_{p})-v_{i_{p}
}^{*}(t_{p})=0$ such that the safety constraint (\ref{Safety}) is satisfied at
$t_{p}$, then $z_{i,i_{p}}(t)>\varphi v_{i}(t)+\delta_{0},\forall t\in\lbrack
t_{i}^{0},t_{i}^{M}]$.
\end{theorem}

\begin{theorem}
\label{thm:merging_general} \citet{Wei2019CDC} Let $i-1>i_{p}$. Under
Assumption \ref{asp:constant}, if $\exists\varepsilon\in(0,1]$ such that
$\varepsilon v_{i}^{0}\leq v_{i-1}^{0}$ and $t_{i}^{0}-t_{i-1}^{0}\geq
\frac{\varphi}{\varepsilon}+\frac{\delta_{0}}{\varepsilon v_{i}^{0}}
+\frac{3L(1-\varepsilon)}{v_{i-1}^{0}+2v_{i-1}^{*}(t_{i-1}^{M})}$, then, under
optimal control (\ref{Optimal_u}) for both $i$ and $i-1$, the safe merging
constraint (\ref{SafeMerging}) is satisfied.
\end{theorem}

Finally, the next result provides conditions such that the speed constraint in
(\ref{VehicleConstraints}) will be satisfied within the CZ:

\begin{theorem}
\label{thm:speedexmp} \citet{Wei2019CDC} If $v_{i}^{0}\leq v_{0},\forall i\in
S(t)$ for $v_{0}\in\lbrack v_{\min} ,v_{\max})$, $\beta>0$ and under optimal
control (\ref{Optimal_u}), then for any $L\leq L_{\max}$, the speed
limitations in (\ref{VehicleConstraints}) are satisfied $\forall t\in\lbrack
t_{i} ^{0},t_{i}^{M}],\forall i\in S(t)$. Where
\[
L_{\max}=\sqrt{\frac{8v_{\max}^{4}-6v_{\max}^{2}v_{0}^{2}-2v_{\max}v_{0}^{3}
}{9\beta}}
\]

\end{theorem}

Note that all conditions in Theorems \ref{thm:safety_general}%
-\ref{thm:speedexmp} are based on the initial conditions $v_{i}^{0},t_{i}^{0}$
of CAV $i\in S(t)$ and information from other CAVs ahead of $i$. Although the
conditions in Theorem. \ref{thm:speedexmp} pertain to all CAVs, it can also be
easily applied to each individual CAV $i\in S(t)$. The case of control
constraints being active is addressed in the following remark.

\begin{remark}
\label{rem:control} If the conditions in Theorems \ref{thm:safety_general}%
-\ref{thm:speedexmp} are satisfied for CAV $i\in S(t)$, but the control
constraint in (\ref{VehicleConstraints}) is initially violated at $u_{max}$
(since we have that $a_{i}<0$ $(\beta\neq0)$ and $u_{i}(t_{i}^{M})=0$ when $i$
is under unconstrained OC (\ref{Optimal_u}) \citet{Wei2019ACC}), then the
safety constraint (\ref{Safety}), the safe merging constraint
(\ref{SafeMerging}) and the speed constraint in (\ref{VehicleConstraints}) are
all satisifed when we first apply $u_{max}$ starting at $t_{i}^{0}$ followed
by an unconstrained OC. This is obvious since the $u_{max}$-constrained OC has
lower speed compared with the unconstrained OC (\ref{Optimal_u}). The
derivation of the unconstrained OC after the $u_{max}$-constrained arc is easy
and time efficient (similar to (\ref{Optimal_u})).
\end{remark}

Once we confirm that a CAV $i\in S(t)$ meets all conditions in Theorems
\ref{thm:safety_general}-\ref{thm:speedexmp} (the control constraint violation
case is discussed in Remark \ref{rem:control} and also viewed as an
unconstrained OC), we can directly apply the unconstrained control
(\ref{Optimal_u}) to CAV $i$. Considering the noisy dynamics
(\ref{VehicleDynamicsNoise}), we wish to find a controller that tracks both
the optimal speed (\ref{Optimal_v}) and position (\ref{Optimal_x}) since the
safety constraint (\ref{Safety}) and the safe merging constraint
(\ref{SafeMerging}) both depend on the speed and position. We use the position
and speed exponential feedback control forms in (\ref{eqn:track}%
)-(\ref{eqn:trackbk}).

Extensive simulation results (see \citet{Wei2019CDC}) have shown that the ratio
of CAVs that satisfy the conditions in Theorems \ref{thm:safety_general}%
-\ref{thm:speedexmp} is large under normal (not exceedingly high) traffic
conditions. Still, when these conditions are not satisfied for some CAV $i\in
S(t)$, we can use the OCBF method to account for these constraints as shown in
the sequel.

\subsection{OCBF for the Merging Problem}

\label{sec:ocbfm} Suppose that an unconstrained OC solution is available for
the objective (\ref{eqn:energyobj}), obtained through (\ref{Optimal_u}%
)-(\ref{Optimal_x}). Our goal here is to determine a controller for those CAVs
that do not satisfy the conditions in Theorems \ref{thm:safety_general}%
-\ref{thm:speedexmp}. This is achieved by combining the unconstrained OC
solution with a CBF-based controller leading to an OCBF controller whose goal
is to track the former as closely as possible.

First, we aim to track the optimal speed $v_{i}^{\ast}(t)$ obtained through
(\ref{Optimal_u})-(\ref{Optimal_x}). In particular, we define a controller
aiming to drive $v_{i}(t)$ to $v_{ref}(t)$ using the form (\ref{StateTrack})
or (\ref{StateTrackAno}). Using either form of $v_{ref}(t)$, we can now
proceed as in (\ref{LyapunovTrack}) and define an output $y_{i}(t):=v_{i}%
(t)-v_{ref}(t)$ and a CLF $V(y_{i}(t))=y_{i}^{2}(t)$. The control should
satisfy the CLF constraint (\ref{LyapunovTrack}).

Second, we deal with the safety and vehicle limitation constraints
(Constraints 1,3) using HOCBFs to map them from the state $\bm x_{i}(t)$ to
the control input $u_{i}(t)$. In particular, define CBFs $b_{i,q}(\bm
x_{i}(t)),$ $q\in\{1,2,3\}$ where $b_{i,1}(\bm x_{i}(t))=v_{max}-v_{i}(t),$
$b_{i,2}(\bm x_{i}(t))=v_{i}(t)-v_{min},$ $b_{i,3}(\bm x_{i}(t))=z_{i,i_{p}%
}(t)-\varphi v_{i}(t)-\delta_{0}$. The relative degree of each $b_{i,q},$
$q\in\{1,2,3\}$ is 1. Therefore, in Definition \ref{def:hocbf}, we choose a
HOCBF with $m=1$. Any control should satisfy the HOCBF constraints
(\ref{eqn:safety0}) and (\ref{eqn:safety1}). Note that $u_{i}(t)\in\lbrack
u_{\min},u_{\max}]$ is already a constraint on the control input, hence, we do
not need to use a HOCBF for it.

Finally, the safe merging constraint (\ref{SafeMerging}) ensures that there
are no collisions when CAVs from different lanes arrive at the merging point
$M$. It is only imposed at $t_{1}^{M}$ and does not apply to all $t\in\lbrack
t_{i}^{0},t_{i}^{M})$. For example, vehicles 4 and 3 in Fig. \ref{fig:merging}
are not constrained before they arrive at the merging point $M$, but have to
satisfy (\ref{SafeMerging}) at $M$. In order to use a HOCBF approach, we need
a version of (\ref{SafeMerging}) that is continuous in time when $i-1>i_{p}$.
Vehicles $i$ and $i-1$ both arrive randomly at $O$ or $O^{\prime}$, and the
minimum distance along the lane $z_{i,i-1}(t_{i}^{0})$ between vehicle $i$ and
$i-1$ is 0, i.e., these two CAVs are allowed to arrive at the origin $O$ or
$O^{\prime}$ at the same time. The coordinator FIFO queue preserves the
arrival order of $i$ and $i-1$ at $O$ or $O^{\prime}$ at the merging point
$M$. When vehicles $i$ and $i-1$ arrive at $M$, they will merge into the same
lane. Therefore, the distance between $i$ and $i-1$ must be greater than or
equal to $\varphi v_{i}(t_{i}^{M})+\delta_{0}$, which is in the form of
(\ref{SafeMerging}). However, we have considerable freedom in choosing the
reaction time $\varphi$ from (\ref{SafeMerging}) for vehicle $i$ ($i-1>i_{p}$)
\ for all $t\in(t_{i}^{0},t_{i}^{M})$. In the following, we provide a
definition for the allowed variation of $\varphi$:

\begin{definition}
The reaction time $\varphi$ for vehicle $i$ ($i-1>i_{p}$) is a strictly
increasing function $\Phi:\mathbb{R}\rightarrow\mathbb{R}$ that satisfies the
initial condition $\Phi(x_{i}(t_{i}^{0}))=-\frac{\delta_{0}}{v_{i}^{0}}$ and
final condition $\Phi(x_{i}(t_{i}^{M}))=\varphi$.
\end{definition}

As an example, in Fig. \ref{fig:merging} where $x_{i}(t_{i}^{0})=0$ and
$x_{i}(t_{i}^{M})=L$, we have $\Phi(x_{i}(t))=\frac{\varphi x_{i}(t)}{L}$ if
$\delta_{0}=0$. The lower bound of the distance from (\ref{SafeMerging})
becomes greater as vehicle $i$ approaches the merging point $M$ such that
there is adequate space for the vehicle in the merging lane to join the main
lane. Therefore, a continuous version of the constraint from
(\ref{SafeMerging}) on $i$ for $i-1>i_{p}$ in the control zone is:
\begin{equation}
z_{i,i-1}(t)\geq\Phi(x_{i}(t))v_{i}(t)+\delta_{0},\text{ \ \ }\forall
t\in\lbrack t_{i}^{0},t_{i}^{M}]. \label{eqn:safemerging}%
\end{equation}
The relative degree of (\ref{eqn:safemerging}) is 1. To enforce safe merging,
we employ a HOCBF that is similar to the ones used for safety
(\ref{eqn:safety0}).

\subsubsection{OCBF controller}

\begin{figure*}[htbp]
	\centering	
	\subfigure[Tracking position error comparison.]{
		\begin{minipage}[t]{0.3\linewidth}
			\centering
			\includegraphics[scale=0.4]{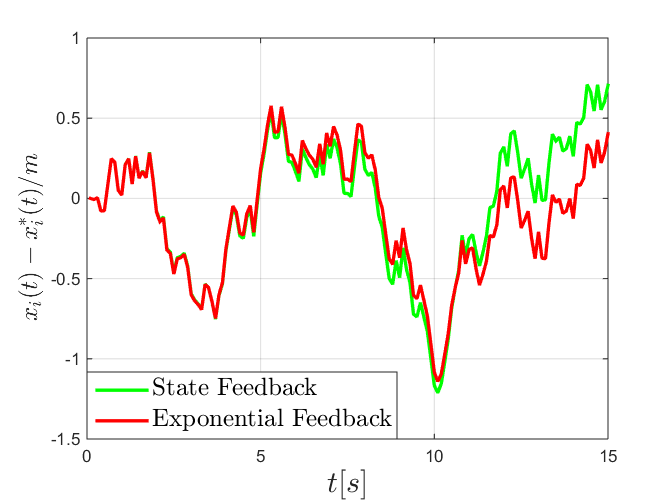} 
			\label{pos_error}%
		\end{minipage}%
	}	
	\subfigure[Tracking speed error comparison.]{
		\begin{minipage}[t]{0.3\linewidth}
			\centering
			\includegraphics[scale=0.4]{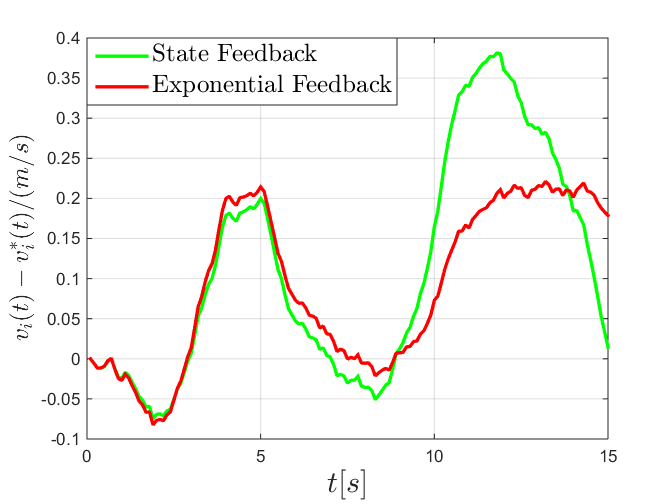} 
			\label{speed_error}%
		\end{minipage}%
	}	
	\subfigure[Control profile comparison.]{
		\begin{minipage}[t]{0.3\linewidth}
			\centering
			\includegraphics[scale=0.4]{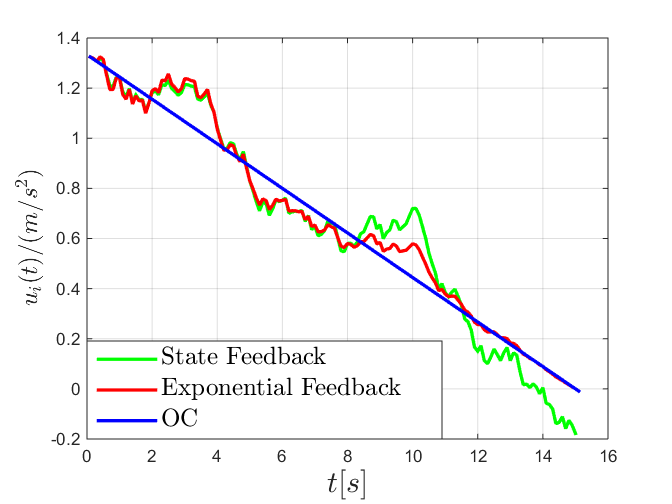} 
			\label{control_error}%
		\end{minipage}%
	}	
	\centering
	\caption{Tracking performance comparison with vehicle noise between the state feedback control (\ref{eqn:trackbk}) and the exponential feedback control (\ref{eqn:track}) {with vehicle limitations (\ref{VehicleConstraints})}.}
	\vspace{-6mm}
\end{figure*}

Along the lines of Sec. \ref{sec:scoc}, we now seek a control input
$u_{i}(t)$ in the HOCBF method which tracks the unconstrained optimal control
$u_{i}^{\ast}(t)$ through a HOCBF controller aiming to drive $u_{i}(t)$ to
$u_{ref}(t)$ defined by (\ref{eqn:track2}) or (\ref{eqn:track}).

Following the OCBF approach in Sec. \ref{sec:scoc}, we apply
(\ref{eqn:objT}) and consider the objective function: {\small
\begin{equation}
J_{i}(u_{i}(t),\delta_{i}(t))\!=\!\int_{t_{i}^{0}}^{t_{i}^{M}}\!\left(
\beta\delta_{i}^{2}(t)\!+\!\frac{1}{2}(u_{i}(t)\!-\!u_{ref}(t))^{2}\right)
dt, \label{eqn:energyobjtrack}%
\end{equation}
}subject to (\ref{VehicleDynamicsNoise}), the corresponding HOCBF constraints
as (\ref{eqn:safety0}), (\ref{eqn:safety1}), and the CLF constraint
(\ref{LyapunovTrack}), the initial and terminal conditions $x_{i}(t_{i}%
^{0})=0$, $x_{i}(t_{i}^{M})=L$, and given $t_{i}^{0},v_{i}(t_{i}^{0})$. Thus,
we have combined the HOCBF method and the OC solution by using
(\ref{eqn:track2}) or (\ref{eqn:track}) to link the optimal position and
acceleration to $u_{ref}(t)$, and use (\ref{StateTrack}) or
(\ref{StateTrackAno}) in the CLF $(v_{i}(t)-v_{ref}(t))^{2}$ to combine with
(\ref{eqn:energyobjtrack}). The resulting optimal $u_{i}(t)$ in
(\ref{eqn:energyobjtrack}) is the OCBF control.

As in (\ref{QP}), we partition the continuous time interval $[t_{i}^{0}%
,t_{i}^{M}]$ into equal time intervals $\{[t_{i}^{0}+k\Delta t,t_{i}%
^{0}+(k+1)\Delta t)\},$ $k=0,1,2,\dots$ In each interval $[t_{i}^{0}+k\Delta
t,t_{i}^{0}+(k+1)\Delta t)$, we assume the control is constant and find a
solution to the optimization problem (\ref{eqn:energyobjtrack}). Specifically,
at $t=t_{i}^{0}+k\Delta t$ ($k=0,1,2,\dots$), we solve
\begin{equation}
\underset{t=t_{i}^{0}+k\Delta t}{\mathbf{QP:}}\bm u_{i}^{\star}%
(t)=\mathop{\arg\min}_{\bm u_{i}(t)}\frac{1}{2}\bm u_{i}(t)^{T}H\bm
u_{i}(t)+F^{T}\bm u_{i}(t)\label{FeasibleQP}%
\end{equation}%
\[
{\small \bm u_{i}(t)=\left[
\begin{array}
[c]{c}%
u_{i}(t)\\
\delta_{i}(t)
\end{array}
\right]  ,}\text{ \ }{\small H=\left[
\begin{array}
[c]{cc}%
1 & 0\\
0 & \beta
\end{array}
\right]  ,}\text{ \ }{\small F=\left[
\begin{array}
[c]{c}%
-u_{ref}(t)\\
0
\end{array}
\right]  }%
\]
subject to the constraints as (\ref{QPcontr1})-(\ref{QPcontr3}) as they
pertain to the merging problem. After solving (\ref{FeasibleQP}) and get an
optimal control ${u}_{i}^{\star}(t)$, we update (\ref{VehicleDynamicsNoise})
for all $t\in(t_{i}^{0}+k\Delta t,t_{i}^{0}+(k+1)\Delta t)$. As shown in Sec.
\ref{sec:simulation}, the use of only (\ref{StateTrack}) or
(\ref{StateTrackAno}), yields an OCBF control which is Lipschitz continuous,
whereas using both state and control trackings improves performance.

\section{SIMULATION RESULTS}

\label{sec:simulation}

All controllers in this section have been implemented using \textsc{MATLAB}
and we have used the Vissim microscopic multi-model traffic flow simulation
tool as a baseline for the purpose of making comparisons between our
controllers and human-driven vehicles adopting standard car-following models
used in Vissim. We used \textsc{quadprog} for solving QPs of the form
(\ref{eqn:energyobjtrack}) or (\ref{eqn:ocbfu}) and \textsc{ode45} to
integrate the vehicle dynamics.

Referring to Fig. \ref{fig:merging}, CAVs arrive according to Poisson
processes with arrival rates that we allow to vary in our simulation examples.
The initial speed $v_{i}(t_{i}^{0})$ is also randomly generated with uniform
distribution in $[15m/s,20m/s]$ at the origins $O$ and $O^{\prime}$,
respectively. The parameters for (\ref{eqn:energyobjtrack}) or
(\ref{eqn:ocbfu}) and (\ref{VehicleDynamicsNoise}) are: $L=400m,\varphi=1.8s,$
$\delta_{0}=0m,u_{max}=3.924m/s^{2},$ $u_{min}=-3.924m/s^{2},$ $v_{max}%
=30m/s,v_{min}=0m/s,$ $\beta=1,$ $\epsilon=10,$ $\Delta t=0.1s,$ $c=1$, and
we consider uniformly distributed noise processes (in [-2, 2] for
$w_{i,1}(t)$ and in [-0.2, 0.2] for $w_{i,2}(t)$) for all simulations. {The value of $\Delta t$ is chosen as small as possible, depending on computational resources available, in order to address the inter-sampling effect on the HOCBFs and maintain a guaranteed satisfaction of all constraints.}

\begin{figure*}[htbp]
	\centering
	
	\subfigure[Controls with only
	speed tracking (\ref{StateTrack}) or (\ref{StateTrackAno}).]{
		\begin{minipage}[t]{0.32\linewidth}
			\centering
			\includegraphics[scale=0.323]{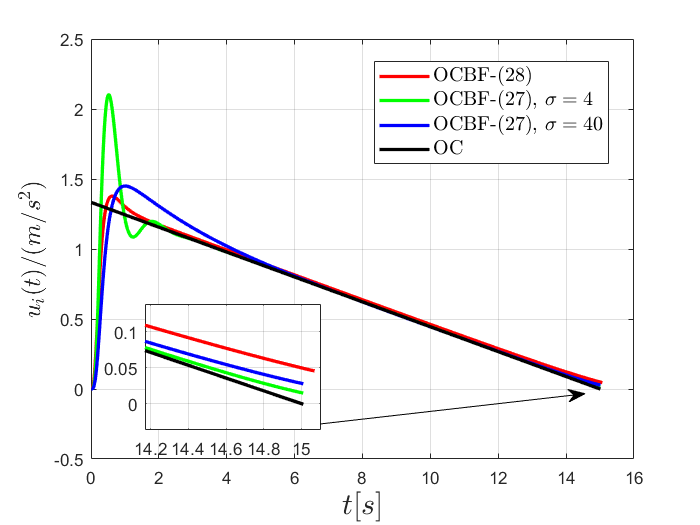} 
			\label{speedonly}%
		\end{minipage}%
	}	
	\subfigure[Controls with both (\ref{StateTrack}) and (\ref{eqn:track2}) under different noise levels. ]{
		\begin{minipage}[t]{0.32\linewidth}
			\centering
			\includegraphics[scale=0.314]{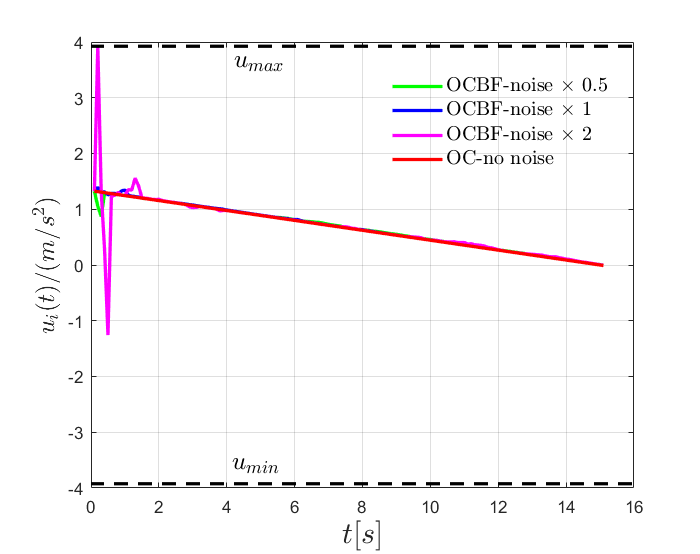} 
			\label{ocbf_noises}%
		\end{minipage}%
	}	
	\subfigure[Controls with (\ref{StateTrackAno}) and  (\ref{eqn:track}), $\sigma \!=\! 40$ under different noise levels.]{
		\begin{minipage}[t]{0.32\linewidth}
			\centering
			\includegraphics[scale=0.328]{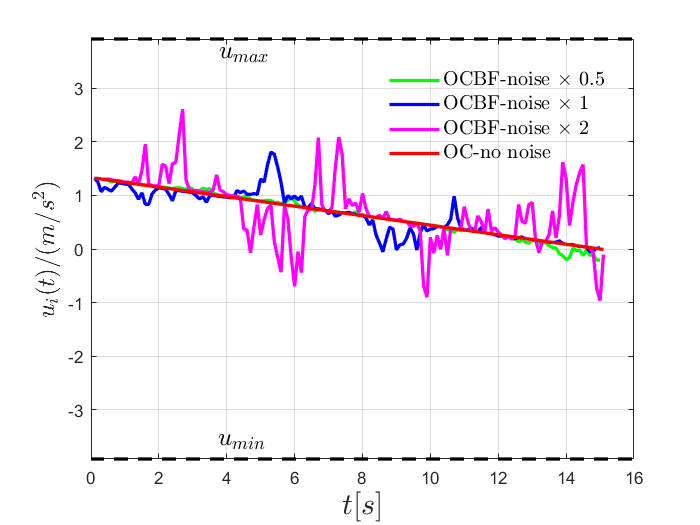} 
			\label{control_noi_ano}%
		\end{minipage}%
	}
	
	\centering
	\caption{OCBF implementation examples under different tracking equations and noise levels { with vehicle limitations (\ref{VehicleConstraints})}.}
\end{figure*}

\textbf{1. Position and speed feedback tracking implementation example.
}First, we provide a simple example of the tracking control implementation for
a single vehicle which considers (\ref{eqn:energyobjtrack}) as the objective
function and employs the unconstrained optimal control (\ref{Optimal_u}).
Although we do not consider the vehicle noise, there is still discretization
($\Delta t=0.1s$) error in the implementation. The initial parameters are
$t_{i}^{0}=0s$, $v_{i}^{0}=20m/s$, $\alpha=0.26$. We first consider the
comparison between exponential feedback control (\ref{eqn:track}) and
directly applied unconstrained control (\ref{Optimal_u}), as shown in terms of
average tracking errors in Table \ref{tab:uncons}. We can see that the
feedback control (\ref{eqn:track}) can significantly improve both
average tracking errors. The tracking errors decrease as $\sigma_{1}%
,\sigma_{2}$ decrease, consistent with the argument after
(\ref{eqn:track}) that we wish to make $\sigma_{1}<\sigma_{2}$, as shown
from the 3rd and 5th columns in Table \ref{tab:uncons}. \begin{table}[ptb]
	\caption{Average tracking error comparison without vehicle noise}%
	\label{tab:uncons}
	\par
	\begin{center}
		\begin{tabular}
			[c]{|c||c||c||c||c|}\hline
			Items & $u^{*}(t)$ (\ref{Optimal_u}) & \multicolumn{3}{|c|}{Feedback control
				(\ref{eqn:track})}\\\hline
			$\sigma_{1},\sigma_{2}$ &  & 4, 12 & 6, 16 & 12, 4\\\hline\hline
			$\frac{1}{2}u_{i}^{2}(t)$ & {\color{red} 4.4000} & 4.4396 & 4.4366 &
			4.4318\\\hline
			Pos. err. & -0.1678 & {\color{red} -0.0280} & -0.0452 & -0.0577\\\hline
			Spd. err. & -0.0333 & {\color{red}-0.0037} & -0.0059 & -0.0095\\\hline
		\end{tabular}
	\end{center}
	\par
\end{table}

Then, under the same randomly generated noise $w_{i,1}(t)\in\lbrack
2m/s,-2m/s]$ and $w_{i,2}(t)\in\lbrack-0.1m/s^{2},0.1m/s^{2}]$, we compare the
tracking performance between the state feedback control
(\ref{eqn:trackbk}) ($k_{1}=0.25,k_{2}=0.1$) and the exponential
feedback control (\ref{eqn:track}) ($\sigma_{1}=4,\sigma_{2}=10$, the
same coefficients as in (\ref{eqn:trackbk})), as shown in Fig.
\ref{pos_error}-\ref{control_error}. We can see that the exponential feedback
control (\ref{eqn:track}) can perform almost the same when the control
$u_{i}^{\ast}(t)$ is large and outperforms the state feedback control
(\ref{eqn:trackbk}) as the optimal control become smaller. The control
input in the exponential feedback control input (\ref{eqn:track}) varies
less than the state feedback control (\ref{eqn:trackbk}), as shown in Fig.
\ref{control_error}.

\textbf{2. OCBF implementation example. }Next, we provide a simple example of
the OCBF controller implementation for a single vehicle which considers
(\ref{eqn:energyobjtrack}) as the objective function. The initial parameters
are the same as the last example. If we only apply
(\ref{StateTrack}) or (\ref{StateTrackAno}), set $u_{ref}(t)=0$ and assume no
noise, then we obtain the control profiles shown in Fig. \ref{speedonly}. The
speed reference form (\ref{StateTrackAno}) tends to achieve a closer track of
the OC control (black curve) compared to the form (\ref{StateTrack}) at the
expense of larger over-shot; as a result, performnace is worse as shown in
Table \ref{tab:sing} (values in red are the best).

If we apply both (\ref{StateTrack}) and (\ref{eqn:track2}) without noise, we
obtain the control profiles shown in where the OCBF
controller's performance is virtually indistinguishable from that of the OC
control, as shown in Table \ref{tab:sing}.

With noise added (based on a uniform distribution in [-2, 2] for $w_{i,1}(t)$ and in [-0.2, 0.2] for $w_{i,2}(t)$), we show the control profiles under different noise levels in
Fig. \ref{ocbf_noises} with (\ref{StateTrack}) and (\ref{eqn:track2}); and
in Fig. \ref{control_noi_ano} with (\ref{StateTrackAno}) and
(\ref{eqn:track}). Constraints 1-3 may be temporarily violated but will
be forced to be satisfied again in finite time through constraint
(\ref{eqn:Lie}). The speed and control tracking forms (\ref{StateTrack}) and
(\ref{eqn:track2}) perform better than (\ref{StateTrackAno}) and
(\ref{eqn:track}) as noise increases.

{\small
	\begin{table}[ptb]
		\caption{Objective function comparison without noise}
		\label{tab:sing}
		\begin{center}%
			\begin{tabular}
				[c]{|c||c||c||c||c||c|}\hline
				Items & OC & \multicolumn{4}{|c|}{OCBF}\\\hline
				Track &  & (\ref{StateTrack}) & (\ref{StateTrackAno}) & (\ref{StateTrackAno}) & (\ref{StateTrack}), (\ref{eqn:track2})\\\hline
				$\sigma$ &  &  & 4 & 40 & \\\hline\hline
				time (s) & {\color{red} 15.01} & 15.07 & 15.01 & 15.01 & 
				15.01\\\hline
				$\frac{1}{2}u_{i}^{2}(t)$ & 4.44 & {\color{red} 4.41} & 4.6962 & 4.66 &
				4.44\\\hline
				objective & {\color{red} 33.33} & 33.43 & 33.52 & 33.50 &
				33.34\\\hline
			\end{tabular}
		\end{center}
	\end{table}
}

\textbf{3. Comparison of OC control from \citet{Wei2019ACC}, CBF control from
	\citet{Wei2019}, and OCBF control in this paper.} Consider the merging
problem with the simple objective function (\ref{eqn:energyobj}) for which we
can easily get unconstrained optimal solutions. Then, we employ the CBF method
and the OCBF technique (with (\ref{StateTrack}) and (\ref{eqn:track2})) introduced in Sec. \ref{sec:ocbfm}. Simulation results
under four different trade-off parameters are shown in Table
\ref{table:comp_OCBF_OC}. We can see that the OCBF method achieves comparable
results to OC, even in the presence of noise. 

The computation time in MATLAB with the OCBF method for each $i$ at each step
is less than 0.01s (Intel(R) Core(TM) i7-8700 CPU @ 3.2GHz$\times 2$), while the
OC method takes between $1s$ and $30s$ for each CAV, depending on whether the
constraints are active or not.

\begin{table}[ptb]
	\caption{Comparison (data in average) of OC, CBF and OCBF (with noise)}
	\par
	\begin{center}%
		\begin{tabular}
			[c]{|c||c|c|c|c|c|}\hline
			Method & $\alpha$ & Noi. & Time($s$) & $\frac{1}{2}u_{i}^{2}(t)$ &
			Obj.\\\hline\hline
			CBF & N/A & no & 14.6978 & 26.9178 & N/A\\\hline\hline
			OC & \multirow{3}*{$0.01$} & no & {\color{red} $\bm {25.4291}$} & {\color{red}
				$\bm {0.1725}$} & {\color{red} $\bm {2.1288}$}\\\cline{1-1}\cline{3-6}%
			\multirow{2}*{OCBF} &  & no & 25.6879 & 1.0582 & 3.0256\\\cline{3-6}
			&  & yes & 25.7494 & 2.2373 & 4.1976\\\hline\hline
			OC & \multirow{3}*{$0.25$} & no & {\color{red} $\bm {17.0472}$} & {\color{red}
				$\bm {4.9069}$} & {\color{red} $\bm {36.4909}$}\\\cline{1-1}\cline{3-6}%
			\multirow{2}*{OCBF} &  & no & 17.1176 & 5.5569 & 37.1139\\\cline{3-6}
			&  & yes & 17.1396 & 6.8959 & 38.1605\\\hline\hline
			OC & \multirow{3}*{$0.40$} & no & {\color{red} $\bm {15.1713}$} & {\color{red}
				$\bm {10.6508}$} & {\color{red} $\bm {53.1120}$}\\\cline{1-1}\cline{3-6}%
			\multirow{2}*{OCBF} &  & no & 15.2286 & 11.3629 & 53.7157\\\cline{3-6}
			&  & yes & 15.2527 & 12.7671 & 54.6325\\\hline
			OC & \multirow{3}*{$0.60$} & no & {\color{red} $\bm {13.1035}$} & {\color{red}
				$\bm {24.4079}$} & {\color{red} $\bm {70.2922}$}\\\cline{1-1}\cline{3-6}%
			\multirow{2}*{OCBF} &  & no & 13.1560 & 25.2468 & 70.8720\\\cline{3-6}
			&  & yes & 13.1692 & 26.6534 & 71.4938\\\hline
		\end{tabular}
	\end{center}
	\par
	\label{table:comp_OCBF_OC}%
\end{table}

\begin{figure}[ptbh]
	\centering
	\includegraphics[scale=0.5]{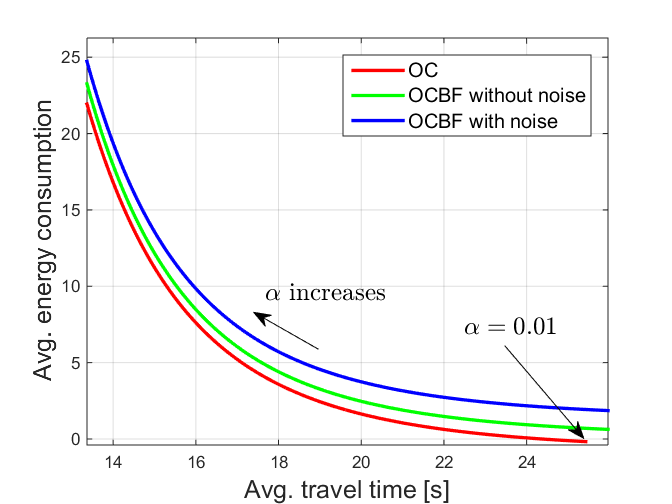} \caption{Travel time and energy
		consumption as the factor $\alpha$ changes.}%
	\label{fig:time_energy}%
\end{figure}

\begin{figure}[ptbh]
	\centering
	\includegraphics[scale=0.5]{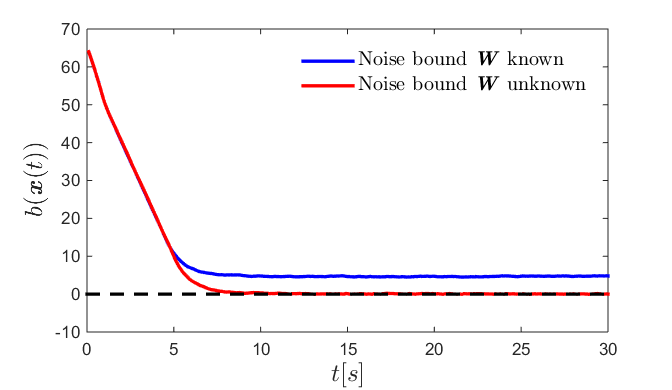} \caption{Barrier function $b(\bm x)$ under noise $w_{i,1}(t)\in[-4,4]m/s, w_{i,2}(t)\in[-0.4,0.4]m/s^2$. $b(\bm x)\geq 0$ denotes the satisfaction of the safety constraint (\ref{Safety}).}%
	\label{fig:bf}%
\end{figure}

We also show in Fig. \ref{fig:time_energy} how the travel time and energy
consumption vary as the weight factor $\alpha$ in (\ref{eqn:energyobja}) changes. The significance of Fig.
\ref{fig:time_energy} is to show how well the OCBF can match the optimal
performance obtained through OC. {Examples of the barrier function profiles for the safety constraint (\ref{Safety}) under known and unknown noise bound $\bm W$ are shown in Fig. \ref{fig:bf}. If $\bm W$ is known, the safety constraint (\ref{Safety}) is guaranteed with some conservativeness; Otherwise, the safety constraint (\ref{Safety}) is satisfied most of the time without conservativeness.}

\textbf{4. Comparison of CBF control from \citet{Wei2019}, CBF control with objective (\ref{eqn:fuelobj}) in this
	paper, and human-driven vehicles through Vissim.} {This simulation refers to \ref{sec:CBF} for the case that the objective function is too complex to get explicit optimal solutions.}
We consider the objective function (\ref{eqn:fuelobj}) which is too
complex to allow the derivation of an OC solution. Thus, we solve
(\ref{eqn:fuelobj}) through the sequence of QPs (\ref{eqn:ocbfu}) and select a value $\beta=0.2$ in (\ref{eqn:ocbfu}) through trial and error to best match the
performance in Vissim. We vary the relative traffic arrival rates of the main
and merging lane and show our results in Tables \ref{table:comp_CBF_Vissim}, \ref{table:comp_CBF_Vissim_main}, \ref{table:comp_CBF_Vissim_merg}.

In Tables \ref{table:comp_CBF_Vissim} and
\ref{table:comp_CBF_Vissim_main}, note that both CBF methods outperform human-driven vehicles modeled though Vissim. We also observe that the CBF method
developed in this paper using (\ref{eqn:fuelobj}) is vastly superior to that
of \citet{Wei2019} in the energy component with little loss in travel time
performance. We also note that without any control (as in Vissim), the main
lane vehicles have priority over the merging lane and the merging lane
vehicles may even stop before the merging point. Thus, there is heavy
congestion in the merging lane when the ratio between the main lane and
merging lane arrival rates is 1:3.

We observe in Table \ref{table:comp_CBF_Vissim_merg} that the energy
consumption of vehicles in Vissim is significantly lower compared to the CBF
methods. This is due to the fact that the merging lane vehicles frequently
stop before the merging point $M$, thus having low speeds when passing over
$M$. In order to achieve a fair comparison, we consider a longer time horizon
over which we measure fuel consumption and travel time. This is accomplished by
extending the trip of each vehicle for an additional length $L$ beyond the
merging point $M$, as shown in Table \ref{table:comp_CBF_Vissim_merg2L}. As
expected, the overall energy performance under CBF control is now significantly better
(by about 37\%) than that of human-driven vehicles. 

\begin{table}[ptb]
	\caption{
		Main lane arrival rate : Merging lane arrival rate = 1:1}
	\par
	\begin{center}%
		\begin{tabular}
			[c]{|c||c||c||c|}\hline
			Items & CBF-(\ref{eqn:energyobj}) & CBF-(\ref{eqn:fuelobj}) &
			Vissim\\\hline
			Ave. time($s$) & {\color{red} $\bm {14.6978}$} & 18.1549 & 25.0813\\\hline
			Main time($s$) & {\color{red} $\bm {14.7000}$} & 18.1717 & 17.9935\\\hline
			Merg. time($s$) & {\color{red}$\bm{14.6956}$} & 18.1378 & 32.3267\\\hline
			Ave. fuel($mL$) & 57.9532 & {\color{red}$\bm{30.9813}$} & 36.9954\\\hline
			Main fuel($mL$) & 57.7028 & {\color{red}$\bm{30.8856}$} & 42.6925\\\hline
			Merg. fuel($mL$) & 58.2092 & {\color{red}$\bm{31.0791}$} & 31.1717\\\hline
		\end{tabular}
	\end{center}
	\par
	\label{table:comp_CBF_Vissim}%
\end{table}

\begin{table}[ptb]
	\caption{ Main lane arrival rate : Merging lane arrival rate = 3:1}
	\par
	\begin{center}%
		\begin{tabular}
			[c]{|c||c||c||c|}\hline
			Items & CBF-(\ref{eqn:energyobj}) & CBF-(\ref{eqn:fuelobj}) &
			Vissim\\\hline
			Ave. time($s$) & {\color{red} $\bm {14.6578}$} & 18.1189 & 23.9300\\\hline
			Main time($s$) & {\color{red} $\bm {14.6794}$} & 18.1413 & 18.3476\\\hline
			Merg. time($s$) & {\color{red}$\bm{14.6074}$} & 18.0667 & 36.9556\\\hline
			Ave. fuel($mL$) & 60.2624 & {\color{red}$\bm{31.9754}$} & 39.8587\\\hline
			Main fuel($mL$) & 61.0934 & {\color{red}$\bm{32.7556}$} & 42.8554\\\hline
			Merg. fuel($mL$) & 58.3235 & {\color{red}$\bm{30.1549}$} & 32.8666\\\hline
		\end{tabular}
	\end{center}
	\par
	\label{table:comp_CBF_Vissim_main}%
\end{table}

\begin{table}[ptb]
	\caption{Main lane arrival rate : Merging lane arrival rate = 1:3}
	\par
	\begin{center}%
		\begin{tabular}
			[c]{|c||c||c||c|}\hline
			Items & CBF-(\ref{eqn:energyobj}) & CBF-(\ref{eqn:fuelobj}) &
			Vissim\\\hline
			Ave. time($s$) & {\color{red} $\bm {14.6000}$} & 18.0093 & 29.2035\\\hline
			Main time($s$) & {\color{red} $\bm {14.7133}$} & 18.1133 & 17.8667\\\hline
			Merg. time($s$) & {\color{red}$\bm{14.5761}$} & 17.9873 & 31.5986\\\hline
			Ave. fuel($mL$) & 61.1607 & 33.4848 & {\color{red}$\bm{30.5212}$}\\\hline
			Main fuel($mL$) & 57.3805 & {\color{red}$\bm{30.9263}$} & 46.5004\\\hline
			Merg. fuel($mL$) & 61.9593 & 34.0253 & {\color{red}$\bm{27.1454}$}\\\hline
		\end{tabular}
	\end{center}
	\par
	\label{table:comp_CBF_Vissim_merg}%
\end{table}

\begin{table}[ptb]
	\caption{  Rate = 1:3, adding a lane of length $L$ after the merging point.}
	\par
	\begin{center}%
		\begin{tabular}
			[c]{|c||c||c||c|}\hline
			Items & CBF-(\ref{eqn:energyobj}) & CBF-(\ref{eqn:fuelobj}) &
			Vissim\\\hline
			Ave. time($s$) & {\color{red} $\bm {28.7975}$} & 36.3076 & 50.9987\\\hline
			Main time($s$) & {\color{red} $\bm {28.9857}$} & 36.3786 & 38.8643\\\hline
			Merg. time($s$) & {\color{red}$\bm{28.7569}$} & 36.2923 & 53.6123\\\hline
			Ave. fuel($mL$) & 88.2784 & {\color{red}$\bm{51.6414}$} & 81.6633\\\hline
			Main fuel($mL$) & 86.6246 & {\color{red}$\bm{48.7578}$} & 77.8110\\\hline
			Merg. fuel($mL$) & 88.6347 & {\color{red}$\bm{52.2625}$} & 82.4930\\\hline
		\end{tabular}
	\end{center}
	\par
	\label{table:comp_CBF_Vissim_merg2L}%
\end{table}

\section{CONCLUSIONS}

\label{sec:conclude}

We have developed a real-time framework that combines optimal trajectories generated through optimal control with the computationally efficient HOCBF method providing safety guarantees.  This allows us to deal with cases where the optimal control solution
becomes computationally costly, as well as to handle the presence of noise in
the system dynamics by exploiting the ability of HOCBFs to
add some robustness to an optimal controller. We applied the proposed framework to the traffic
merging problem for connected and automated vehicles with results showing significant improvement in performance compared with human driven vehicles. An ongoing research challenge is imparting adaptivity to HOCBF-based controllers with respect to a changing environment. Regarding autonomous vehicles (CAVs) in a traffic network, ongoing work is aimed at integrating them with non-CAVs.


\bibliographystyle{elsarticle-harv}
\bibliography{autosam}

\begin{thebibliography}{44}
\expandafter\ifx\csname natexlab\endcsname\relax\def\natexlab#1{#1}\fi
\providecommand{\url}[1]{\texttt{#1}}
\providecommand{\href}[2]{#2}
\providecommand{\path}[1]{#1}
\providecommand{\DOIprefix}{doi:}
\providecommand{\ArXivprefix}{arXiv:}
\providecommand{\URLprefix}{URL: }
\providecommand{\Pubmedprefix}{pmid:}
\providecommand{\doi}[1]{\href{http://dx.doi.org/#1}{\path{#1}}}
\providecommand{\Pubmed}[1]{\href{pmid:#1}{\path{#1}}}
\providecommand{\bibinfo}[2]{#2}
\ifx\xfnm\relax \def\xfnm[#1]{\unskip,\space#1}\fi
\bibitem[{Ames et~al.(2012)Ames, Galloway and Grizzle}]{Aaron2012}
\bibinfo{author}{Ames, A.D.}, \bibinfo{author}{Galloway, K.},
  \bibinfo{author}{Grizzle, J.W.}, \bibinfo{year}{2012}.
\newblock \bibinfo{title}{Control lyapunov functions and hybrid zero dynamics},
  in: \bibinfo{booktitle}{Proc. of 51rd IEEE Conference on Decision and
  Control}, pp. \bibinfo{pages}{6837--6842}.
\bibitem[{Ames et~al.(2017)Ames, Xu, Grizzle and Tabuada}]{Ames2017}
\bibinfo{author}{Ames, A.D.}, \bibinfo{author}{Xu, X.},
  \bibinfo{author}{Grizzle, J.W.}, \bibinfo{author}{Tabuada, P.},
  \bibinfo{year}{2017}.
\newblock \bibinfo{title}{Control barrier function based quadratic programs for
  safety critical systems}.
\newblock \bibinfo{journal}{IEEE Transactions on Automatic Control}
  \bibinfo{volume}{62}, \bibinfo{pages}{3861--3876}.
\bibitem[{Ansari and Murphey(2016)}]{Ansari2016}
\bibinfo{author}{Ansari, A.R.}, \bibinfo{author}{Murphey, T.D.},
  \bibinfo{year}{2016}.
\newblock \bibinfo{title}{Sequential action control: Closed-form optimal
  control for nonlinear and nonsmooth systems}.
\newblock \bibinfo{journal}{IEEE Transactions on Robotics}
  \bibinfo{volume}{32}, \bibinfo{pages}{1196--1214}.
\bibitem[{Aubin(2009)}]{Aubin2009}
\bibinfo{author}{Aubin, J.P.}, \bibinfo{year}{2009}.
\newblock \bibinfo{title}{Viability theory}.
\newblock \bibinfo{publisher}{Springer}.
\bibitem[{Bemporad et~al.(2002)Bemporad, Borrelli and Morari}]{Bemporad2002}
\bibinfo{author}{Bemporad, A.}, \bibinfo{author}{Borrelli, F.},
  \bibinfo{author}{Morari, M.}, \bibinfo{year}{2002}.
\newblock \bibinfo{title}{Model predictive control based on linear programming,
  the explicit solution}.
\newblock \bibinfo{journal}{IEEE transactions on automatic control}
  \bibinfo{volume}{47}, \bibinfo{pages}{1974--1985}.
\bibitem[{Boyd and Vandenberghe(2004)}]{Boyd2004}
\bibinfo{author}{Boyd, S.P.}, \bibinfo{author}{Vandenberghe, L.},
  \bibinfo{year}{2004}.
\newblock \bibinfo{title}{Convex optimization}.
\newblock \bibinfo{publisher}{Cambridge university press},
  \bibinfo{address}{New York}.
\bibitem[{Bryson and Ho(1969)}]{Bryson1969}
\bibinfo{author}{Bryson}, \bibinfo{author}{Ho}, \bibinfo{year}{1969}.
\newblock \bibinfo{title}{Applied Optimal Control}.
\newblock \bibinfo{publisher}{Ginn Blaisdell}, \bibinfo{address}{Waltham, MA}.
\bibitem[{Cao et~al.(2015)Cao, Mukai and Kawabe}]{Cao2015}
\bibinfo{author}{Cao, W.}, \bibinfo{author}{Mukai, M.},
  \bibinfo{author}{Kawabe, T.}, \bibinfo{year}{2015}.
\newblock \bibinfo{title}{Cooperative vehicle path generation during merging
  using model predictive control with real-time optimization}.
\newblock \bibinfo{journal}{Control Engineering Practice} \bibinfo{volume}{34},
  \bibinfo{pages}{98--105}.
\bibitem[{Chitour et~al.(2012)Chitour, Jean and Mason}]{Chitour2012}
\bibinfo{author}{Chitour, Y.}, \bibinfo{author}{Jean, F.},
  \bibinfo{author}{Mason, P.}, \bibinfo{year}{2012}.
\newblock \bibinfo{title}{Optimal control models of goal-oriented human
  locomotion}.
\newblock \bibinfo{journal}{SIAM Journal on Control and Optimization}
  \bibinfo{volume}{50}, \bibinfo{pages}{147--170}.
\bibitem[{Freeman and Kokotovic(1996)}]{Freeman1996}
\bibinfo{author}{Freeman, R.A.}, \bibinfo{author}{Kokotovic, P.V.},
  \bibinfo{year}{1996}.
\newblock \bibinfo{title}{Robust Nonlinear Control Design}.
\newblock \bibinfo{publisher}{Birkhauser}.
\bibitem[{Galloway et~al.(2015)Galloway, Sreenath, Ames and
  Grizzle}]{Galloway2013}
\bibinfo{author}{Galloway, K.}, \bibinfo{author}{Sreenath, K.},
  \bibinfo{author}{Ames, A.D.}, \bibinfo{author}{Grizzle, J.},
  \bibinfo{year}{2015}.
\newblock \bibinfo{title}{Torque saturation in bipedal robotic walking through
  control lyapunov function based quadratic programs}.
\newblock \bibinfo{journal}{IEEE Access} \bibinfo{volume}{50},
  \bibinfo{pages}{323--332}.
\bibitem[{Garcia and Prett(1989)}]{Garcia1989}
\bibinfo{author}{Garcia, C.E.}, \bibinfo{author}{Prett, D.M.},
  \bibinfo{year}{1989}.
\newblock \bibinfo{title}{Model predictive control: theory and practice}.
\newblock \bibinfo{journal}{Automatica} \bibinfo{volume}{25},
  \bibinfo{pages}{335--348}.
\bibitem[{Glotfelter et~al.(2017)Glotfelter, Cortes and
  Egerstedt}]{Glotfelter2017}
\bibinfo{author}{Glotfelter, P.}, \bibinfo{author}{Cortes, J.},
  \bibinfo{author}{Egerstedt, M.}, \bibinfo{year}{2017}.
\newblock \bibinfo{title}{Nonsmooth barrier functions with applications to
  multi-robot systems}.
\newblock \bibinfo{journal}{IEEE control systems letters} \bibinfo{volume}{1},
  \bibinfo{pages}{310--315}.
\bibitem[{Kamal et~al.(2013)Kamal, Mukai, Murata and Kawabe}]{Kamal2013}
\bibinfo{author}{Kamal, M.}, \bibinfo{author}{Mukai, M.},
  \bibinfo{author}{Murata, J.}, \bibinfo{author}{Kawabe, T.},
  \bibinfo{year}{2013}.
\newblock \bibinfo{title}{Model predictive control of vehicles on urban roads
  for improved fuel economy}.
\newblock \bibinfo{journal}{IEEE Transactions on Control Systems Technology}
  \bibinfo{volume}{21}, \bibinfo{pages}{831--841}.
\bibitem[{Khalil(2002)}]{Khalil2002}
\bibinfo{author}{Khalil, H.K.}, \bibinfo{year}{2002}.
\newblock \bibinfo{title}{Nonlinear Systems}.
\newblock \bibinfo{publisher}{Prentice Hall, third edition}.
\bibitem[{Levine and Athans(1966)}]{Levine1966}
\bibinfo{author}{Levine, W.}, \bibinfo{author}{Athans, M.},
  \bibinfo{year}{1966}.
\newblock \bibinfo{title}{On the optimal error regulation of a string of moving
  vehicles}.
\newblock \bibinfo{journal}{IEEE Transactions on Automatic Control}
  \bibinfo{volume}{11}, \bibinfo{pages}{355--361}.
\bibitem[{Lindemann and Dimarogonas(2019)}]{Lindemann2019}
\bibinfo{author}{Lindemann, L.}, \bibinfo{author}{Dimarogonas, D.V.},
  \bibinfo{year}{2019}.
\newblock \bibinfo{title}{Control barrier functions for multi-agent systems
  under conflicting local signal temporal logic tasks}.
\newblock \bibinfo{journal}{IEEE Control Systems Letters} \bibinfo{volume}{3},
  \bibinfo{pages}{757--762}.
\bibitem[{Malikopoulos et~al.(2018)Malikopoulos, Cassandras and
  Zhang}]{Malikopoulos2018}
\bibinfo{author}{Malikopoulos, A.A.}, \bibinfo{author}{Cassandras, C.G.},
  \bibinfo{author}{Zhang, Y.J.}, \bibinfo{year}{2018}.
\newblock \bibinfo{title}{A decentralized energy-optimal control framework for
  connected and automated vehicles at signal-free intersections}.
\newblock \bibinfo{journal}{Automatica} \bibinfo{volume}{2018},
  \bibinfo{pages}{244--256}.
\bibitem[{Mayne(2014)}]{Mayne2014}
\bibinfo{author}{Mayne, D.}, \bibinfo{year}{2014}.
\newblock \bibinfo{title}{Model predictive control: Recent developments and
  future promise}.
\newblock \bibinfo{journal}{Automatica} \bibinfo{volume}{50},
  \bibinfo{pages}{2967--2986}.
\bibitem[{Milanes et~al.(2012)Milanes, Godoy, Villagra and Perez}]{Milanes2012}
\bibinfo{author}{Milanes, V.}, \bibinfo{author}{Godoy, J.},
  \bibinfo{author}{Villagra, J.}, \bibinfo{author}{Perez, J.},
  \bibinfo{year}{2012}.
\newblock \bibinfo{title}{Automated on-ramp merging system for congested
  traffic situations}.
\newblock \bibinfo{journal}{IEEE Transactions on Intelligent Transportation
  Systems} \bibinfo{volume}{12}, \bibinfo{pages}{500--508}.
\bibitem[{Mita et~al.(2001)Mita, Nam and Hyon}]{Mita2001}
\bibinfo{author}{Mita, T.}, \bibinfo{author}{Nam, T.K.}, \bibinfo{author}{Hyon,
  S.H.}, \bibinfo{year}{2001}.
\newblock \bibinfo{title}{Analytical time optimal control solution for a 2-link
  free flying acrobots}, in: \bibinfo{booktitle}{Proc. of IEEE International
  Conference on Robotics and Automation}, pp. \bibinfo{pages}{2741--2746}.
\bibitem[{Mukai et~al.(2017)Mukai, Natori and Fujita}]{Mukai2017}
\bibinfo{author}{Mukai, M.}, \bibinfo{author}{Natori, H.},
  \bibinfo{author}{Fujita, M.}, \bibinfo{year}{2017}.
\newblock \bibinfo{title}{Model predictive control with a mixed integer
  programming for merging path generation on motor way}, in:
  \bibinfo{booktitle}{Proc. IEEE Conference on Control Technology and
  Applications}, \bibinfo{address}{pp. 2214--2219, Mauna Lani}.
\bibitem[{Nguyen and Sreenath(2016)}]{Nguyen2016}
\bibinfo{author}{Nguyen, Q.}, \bibinfo{author}{Sreenath, K.},
  \bibinfo{year}{2016}.
\newblock \bibinfo{title}{Exponential control barrier functions for enforcing
  high relative-degree safety-critical constraints}, in:
  \bibinfo{booktitle}{Proc. of the American Control Conference}, pp.
  \bibinfo{pages}{322--328}.
\bibitem[{Ntousakis et~al.(2016)Ntousakis, Nikolos and
  Papageorgiou}]{Ntousakis2016}
\bibinfo{author}{Ntousakis, I.A.}, \bibinfo{author}{Nikolos, I.K.},
  \bibinfo{author}{Papageorgiou, M.}, \bibinfo{year}{2016}.
\newblock \bibinfo{title}{Optimal vehicle trajectory planning in the context of
  cooperative merging on highways}.
\newblock \bibinfo{journal}{Transportation Research Part C}
  \bibinfo{volume}{71}, \bibinfo{pages}{464--488}.
\bibitem[{Panagou et~al.(2013)Panagou, Stipanovic and Voulgaris}]{Panagou2013}
\bibinfo{author}{Panagou, D.}, \bibinfo{author}{Stipanovic, D.M.},
  \bibinfo{author}{Voulgaris, P.G.}, \bibinfo{year}{2013}.
\newblock \bibinfo{title}{Multi-objective control for multi-agent systems using
  lyapunov-like barrier functions}, in: \bibinfo{booktitle}{Proc. of 52nd IEEE
  Conference on Decision and Control}, \bibinfo{address}{Florence, Italy}. pp.
  \bibinfo{pages}{1478--1483}.
\bibitem[{Prajna et~al.(2007)Prajna, Jadbabaie and Pappas}]{Prajna2007}
\bibinfo{author}{Prajna, S.}, \bibinfo{author}{Jadbabaie, A.},
  \bibinfo{author}{Pappas, G.J.}, \bibinfo{year}{2007}.
\newblock \bibinfo{title}{A framework for worst-case and stochastic safety
  verification using barrier certificates}.
\newblock \bibinfo{journal}{IEEE Transactions on Automatic Control}
  \bibinfo{volume}{52}, \bibinfo{pages}{1415--1428}.
\bibitem[{Raravi et~al.(2007)Raravi, Shingde, Ramamritham and
  Bharadia}]{Raravi2007}
\bibinfo{author}{Raravi, G.}, \bibinfo{author}{Shingde, V.},
  \bibinfo{author}{Ramamritham, K.}, \bibinfo{author}{Bharadia, J.},
  \bibinfo{year}{2007}.
\newblock \bibinfo{title}{Merge algorithms for intelligent vehicles. In:
  Sampath, P., Ramesh, S. (Eds.), Next Generation Design and Verification
  Methodologies for Distributed Embedded Control Systems}.
\newblock \bibinfo{publisher}{Springer}, \bibinfo{address}{Waltham, MA}.
\bibitem[{Rathgeber et~al.(2015)Rathgeber, Winkler, Kang and
  Muller}]{Rathgeber2015}
\bibinfo{author}{Rathgeber, C.}, \bibinfo{author}{Winkler, F.},
  \bibinfo{author}{Kang, X.}, \bibinfo{author}{Muller, S.},
  \bibinfo{year}{2015}.
\newblock \bibinfo{title}{Optimal trajectories for highly automated driving}.
\newblock \bibinfo{journal}{International Journal of Mechanical, Aerospace,
  Industrial, Mechatronic and Manufacturing Engineering} \bibinfo{volume}{9},
  \bibinfo{pages}{946--952}.
\bibitem[{Rios-Torres and Malikopoulos(2017)}]{Torres2015}
\bibinfo{author}{Rios-Torres, J.}, \bibinfo{author}{Malikopoulos, A.},
  \bibinfo{year}{2017}.
\newblock \bibinfo{title}{Automated and cooperative vehicle merging at highway
  on-ramps}.
\newblock \bibinfo{journal}{IEEE Transactions on Intelligent Transportation
  Systems} \bibinfo{volume}{18}, \bibinfo{pages}{780--789}.
\bibitem[{Scarinci and Heydecker(2014)}]{Scarinci2014}
\bibinfo{author}{Scarinci, R.}, \bibinfo{author}{Heydecker, B.},
  \bibinfo{year}{2014}.
\newblock \bibinfo{title}{Control concepts for facilitating motorway on-ramp
  merging using intelligent vehicles}.
\newblock \bibinfo{journal}{Transport Reviews} \bibinfo{volume}{34},
  \bibinfo{pages}{775--797}.
\bibitem[{Schrank et~al.(2015)Schrank, Eisele, Lomax and Bak}]{Schrank2015}
\bibinfo{author}{Schrank, B.}, \bibinfo{author}{Eisele, B.},
  \bibinfo{author}{Lomax, T.}, \bibinfo{author}{Bak, J.}, \bibinfo{year}{2015}.
\newblock \bibinfo{title}{The 2015 urban mobility scorecard}.
\newblock \bibinfo{howpublished}{Texas A\&M Transportation Institute}.
\newblock \URLprefix \url{http://mobility.tamu.edu}.
\bibitem[{Sontag(1983)}]{Sontag1983}
\bibinfo{author}{Sontag, E.}, \bibinfo{year}{1983}.
\newblock \bibinfo{title}{A lyapunov-like stabilization of asymptotic
  controllability}.
\newblock \bibinfo{journal}{SIAM Journal of Control and Optimization}
  \bibinfo{volume}{21}, \bibinfo{pages}{462--471}.
\bibitem[{Tideman et~al.(2007)Tideman, van~der Voort, van Arem and
  Tillema}]{Tideman2007}
\bibinfo{author}{Tideman, M.}, \bibinfo{author}{van~der Voort, M.},
  \bibinfo{author}{van Arem, B.}, \bibinfo{author}{Tillema, F.},
  \bibinfo{year}{2007}.
\newblock \bibinfo{title}{A review of lateral driver support systems}, in:
  \bibinfo{booktitle}{Proc. IEEE Intelligent Transportation Systems
  Conference}, \bibinfo{address}{pp. 992--999, Seatle}.
\bibitem[{Varaiya(1993)}]{Varaiya1993}
\bibinfo{author}{Varaiya, P.}, \bibinfo{year}{1993}.
\newblock \bibinfo{title}{Smart cars on smart roads: problems of control}.
\newblock \bibinfo{journal}{IEEE Transactions on Automatic Control}
  \bibinfo{volume}{38}, \bibinfo{pages}{195--207}.
\bibitem[{Vogel(2003)}]{Vogel2003}
\bibinfo{author}{Vogel, K.}, \bibinfo{year}{2003}.
\newblock \bibinfo{title}{A comparison of headway and time to collision as
  safety indicators}.
\newblock \bibinfo{journal}{Accident Analysis \& Prevention}
  \bibinfo{volume}{35}, \bibinfo{pages}{427--433}.
\bibitem[{Waard et~al.(2009)Waard, Dijksterhuis and Broohuis}]{Waard2009}
\bibinfo{author}{Waard, D.D.}, \bibinfo{author}{Dijksterhuis, C.},
  \bibinfo{author}{Broohuis, K.A.}, \bibinfo{year}{2009}.
\newblock \bibinfo{title}{Merging into heavy motorway traffic by young and
  elderly drivers}.
\newblock \bibinfo{journal}{Accident Analysis and Prevention}
  \bibinfo{volume}{41}, \bibinfo{pages}{588--597}.
\bibitem[{Wieland and Allgower(2007)}]{Wieland2007}
\bibinfo{author}{Wieland, P.}, \bibinfo{author}{Allgower, F.},
  \bibinfo{year}{2007}.
\newblock \bibinfo{title}{Constructive safety using control barrier functions},
  in: \bibinfo{booktitle}{Proc. of 7th IFAC Symposium on Nonlinear Control
  System}.
\bibitem[{Wisniewski and Sloth(2013)}]{Wisniewski2013}
\bibinfo{author}{Wisniewski, R.}, \bibinfo{author}{Sloth, C.},
  \bibinfo{year}{2013}.
\newblock \bibinfo{title}{Converse barrier certificate theorem}, in:
  \bibinfo{booktitle}{Proc. of 52nd IEEE Conference on Decision and Control},
  \bibinfo{address}{Florence, Italy}. pp. \bibinfo{pages}{4713--4718}.
\bibitem[{Xiao and Belta(2019)}]{Xiao2019}
\bibinfo{author}{Xiao, W.}, \bibinfo{author}{Belta, C.}, \bibinfo{year}{2019}.
\newblock \bibinfo{title}{Control barrier functions for systems with high
  relative degree}, in: \bibinfo{booktitle}{Proc. of 58th IEEE Conference on
  Decision and Control}, \bibinfo{address}{Nice, France}. pp.
  \bibinfo{pages}{474--479}.
\bibitem[{Xiao et~al.(2019a)Xiao, Belta and Cassandras}]{Wei2019}
\bibinfo{author}{Xiao, W.}, \bibinfo{author}{Belta, C.},
  \bibinfo{author}{Cassandras, C.G.}, \bibinfo{year}{2019}a.
\newblock \bibinfo{title}{Decentralized merging control in traffic networks: A
  control barrier function approach}, in: \bibinfo{booktitle}{Proc. ACM/IEEE
  International Conference on Cyber-Physical Systems},
  \bibinfo{address}{Montreal, Canada}. pp. \bibinfo{pages}{270--279}.
\bibitem[{Xiao and Cassandras(2019a)}]{Wei2019CDC}
\bibinfo{author}{Xiao, W.}, \bibinfo{author}{Cassandras, C.G.},
  \bibinfo{year}{2019}a.
\newblock \bibinfo{title}{Conditions for improving the computational efficiency
  of decentralized optimal merging controllers for connected and automated
  vehicles}, in: \bibinfo{booktitle}{Proc. of 58th IEEE Conference on Decision
  and Control}, \bibinfo{address}{Nice, France}. pp.
  \bibinfo{pages}{3158--3163}.
\bibitem[{Xiao and Cassandras(2019b)}]{Wei2019ACC}
\bibinfo{author}{Xiao, W.}, \bibinfo{author}{Cassandras, C.G.},
  \bibinfo{year}{2019}b.
\newblock \bibinfo{title}{Decentralized optimal merging control for connected
  and automated vehicles}, in: \bibinfo{booktitle}{Proc. of the American
  Control Conference}, pp. \bibinfo{pages}{3315--3320}.
\bibitem[{Xiao and Cassandras(2020)}]{Wei2020ACC}
\bibinfo{author}{Xiao, W.}, \bibinfo{author}{Cassandras, C.G.},
  \bibinfo{year}{2020}.
\newblock \bibinfo{title}{Decentralized optimal merging control for connected
  and automated vehicles with optimal dynamic resequencing}, in:
  \bibinfo{booktitle}{Proc. of the American Control Conference}, pp.
  \bibinfo{pages}{4090--4095}.
\bibitem[{Xiao et~al.(2019b)Xiao, Cassandras and Belta}]{Wei2019itsc}
\bibinfo{author}{Xiao, W.}, \bibinfo{author}{Cassandras, C.G.},
  \bibinfo{author}{Belta, C.}, \bibinfo{year}{2019}b.
\newblock \bibinfo{title}{Decentralized merging control in traffic networks
  with noisy vehicle dynamics: A joint optimal control and barrier function
  approach}, in: \bibinfo{booktitle}{Proc. IEEE 22nd Intelligent Transportation
  Systems Conference}, \bibinfo{address}{Auckland, New Zealand}. pp.
  \bibinfo{pages}{3162--3167}.

\end{thebibliography}

\appendix
\section{Complex Objectives, Dynamics and Comfort }

\label{sec:CBF}

As shown in \citet{Wei2019}, the HOCBF method allows us to deal with nonlinear
systems and to consider more complex objective functions than
(\ref{eqn:energyobj}). In particular, we consider:
\begin{equation}
\min_{u_{i}(t)}\beta(t_{i}^{M}-t_{i}^{0})+\int_{t_{i}^{0}%
}^{t_{i}^{M}}f_{v}(t)dt, \label{eqn:fuelobj}%
\end{equation}
where $f_{v}(t)$ represents a more detailed realistic energy model replacing
the simple expression $u_{i}^{2}(t)$ commonly used as a surrogate energy
function. As an example, we have adopted in \citet{Wei2019} the following
energy model from \citet{Kamal2013}, which describes fuel consumed per second
as
\begin{equation}
\begin{aligned} f_{v}(t)&=f_{cruise}(t)+f_{accel}(t),\\ f_{cruise}(t)&=\omega_{0}+\omega_{1}v_{i}(t)+\omega_{2}v_{i}^{2}(t)+\omega _{3}v_{i}^{3}(t),\\ f_{accel}(t)&=(r_{0}+r_{1}v_{i}(t)+r_{2}v_{i}^{2}(t))u_{i}(t). \end{aligned} \label{FuelModel}%
\end{equation}
where $\omega_{0}$, $\omega_{1}$, $\omega_{2}$, $\omega_{3}$, $r_{0}$, $r_{1}$
and $r_{2}$ are positive coefficients (typical values are reported in
\citet{Kamal2013}). It is assumed that during braking, i.e., $u_{i}(t)<0$, no
fuel is consumed. Note that (\ref{eqn:fuelobj}) is hard to solve through an OC
analysis as in the previous section. However, in the HOCBF approach this can be
handled numerically.

As for the dynamics of CAVs, the HOCBF method can easily handle nonlinear
dynamics instead of just the linear form in (\ref{VehicleDynamics}). Thus, we
use the vehicle dynamics \citet{Khalil2002}:
\begin{equation}
\underbrace{\left[
	\begin{array}
	[c]{c}%
	\dot{x}_{i}(t)\\
	\dot{v}_{i}(t)
	\end{array}
	\right]  }_{\dot{\bm x}_{i}(t)}=\underbrace{\left[
	\begin{array}
	[c]{c}%
	v_{i}(t)\\
	-\frac{1}{m_{i}}F_{r}(v_{i}(t))
	\end{array}
	\right]  }_{f(\bm x_{i}(t))}+\underbrace{\left[
	\begin{array}
	[c]{c}%
	0\\
	\frac{1}{m_{i}}%
	\end{array}
	\right]  }_{g(\bm x_{i}(t))}u_{i}(t), \label{VehicleModel}%
\end{equation}
where $m_{i}$ denotes the mass of CAV $i$, and $v_{i}(t)$ is its velocity.
$F_{r}(v_{i}(t))$ denotes the resistance force, which is normally expressed
\citet{Khalil2002} as:
\begin{equation}
F_{r}(v_{i}(t))=k_{0}sgn(v_{i}(t))+k_{1}v_{i}(t)+k_{2}v_{i}^{2}(t),
\end{equation}
where $k_{0}>0,k_{1}>0$ and $k_{2}>0$ are scalars determined empirically, and
$sgn$ is the signum function. The first term in $F_{r}(v_{i}(t))$ denotes the
Coulomb friction force, the second term denotes the viscous friction force and
the last term denotes the aerodynamic drag.

In the HOCBF method, we do not explicitly optimize the travel time shown in
(\ref{eqn:fuelobj}). Instead, we use a CLF to drive $v_{i}(t)$ to a desired
speed such that the travel time is optimized. In \citet{Wei2019}, we define an
output $y_{i}(t):=v_{i}(t)-v_{max}$ and choose a CLF $V(y_{i}(t))=y_{i}%
^{2}(t)$. Any control input $u_{i}(t)$ should satisfy, for all $t\in\lbrack
t_{i}^{0},t_{i}^{M}]$,%
\begin{equation}
\begin{aligned} L_fV(y_i(t)) + L_gV(y_i(t))u_i(t) + \epsilon V(y_i(t)) \leq \delta_{i}(t) \end{aligned} \label{LyapunovConstraintold}%
\end{equation}
where $\epsilon>0$ and $\delta_{i}(t)$ is a relaxation variable that makes the
requirement $v_{i}(t)=v_{max}$ to be treated as a soft constraint. Thus, we
seek to achieve Objective 1 indirectly and consider Objective 2 directly,
replacing (\ref{eqn:fuelobj}) by
\begin{equation}
\min_{u_{i}(t),\delta_{i}(t)}\int_{t_{i}^{0}%
}^{t_{i}^{M}}\left(f_{v}(t)+\beta\delta_{i}^{2}(t)\right)dt \label{eqn:ocbfu}%
\end{equation}
subject to the same constraints as in (\ref{eqn:energyobjtrack}) and dynamics
(\ref{VehicleModel}). We use the QP-based method as introduced in the last
subsection to solve (\ref{eqn:ocbfu}). Thus, all CAVs can safely pass over the
merging point $M$ while minimizing $J_{i}(u_{i}(t),\delta_{i}(t))$ within each
time interval, hence jointly minimizing the energy consumption captured by
$f_{v}(t)$ and travel time (indirectly) through the minimization of
$\delta_{i}^{2}$. By adjusting the weight $\beta$ in (\ref{eqn:ocbfu}), we can
trade off between these two objectives.

When comfort is also concerned in the objective, i.e., we also want to
minimize the jerk of each CAV $i$, we can directly incorporate the jerk into
(\ref{eqn:ocbfu}). Noting that $f_{v}(t)$ in (\ref{eqn:ocbfu}) is linear in
$u_{i}(t)$, we wish to formulate a Linear Program (LP) instead of a QP since
the LP tends to be around 30\% more computationally efficient than the QP, as
shown in \citet{Wei2019}. Including the comfort requirement, we have
\begin{equation}
\min_{u_{i}(t),\delta_{i}(t)}\int_{t_{i}%
	^{0}}^{t_{i}^{M}}f_{v}(t)\!+\!\beta_{1}\delta_{i}(t)\!+\!\beta_{2}\left\vert
\frac{u_{i}(t)\!-\!u_{i}^{\ast}(t\!-\!k\Delta t)}{\Delta t}\right\vert dt
\label{eqn:ocbfcomf}%
\end{equation}
where $u_{i}^{\ast}(t-k\Delta t)$ denotes the optimal control from the last
time interval (initially set to 0 at $t_{i}^{0}$), and is known. The
parameters $\beta_{1}>0,$ $\beta_{2}>0$ trade off fuel consumption, travel
time, and comfort. The LP (\ref{eqn:ocbfcomf}) is subject to the same
constraints as the QP (\ref{eqn:ocbfu}).

\end{document}